\documentclass[format=sigconf, screen=true, 10pt]{acmart}

\newif\ifnotnewacm
\notnewacmfalse

\usepackage{color}
\usepackage{cleveref}
\usepackage{tabularx}
\usepackage{xfrac}
\usepackage{amsmath}
\usepackage{caption}
\usepackage{subcaption}
\usepackage{graphicx}
\usepackage{gensymb}
\usepackage{siunitx}
\usepackage{bm}
\usepackage{multicol}
\usepackage{caption}
\usepackage[ruled,vlined]{algorithm2e}
\usepackage{enumitem}
\usepackage{makecell}
\usepackage{colortbl}
\usepackage{xcolor}
\usepackage{libertine}
\usepackage{diagbox}
\usepackage{booktabs}

\newtheorem{proposition}{\textbf{Proposition}}
\newif\ifeg
\egfalse

\acmBooktitle{The 31st Annual International Conference on Mobile Computing and Networking (ACM MOBICOM '25), November 4--8, 2025, Hong Kong, China}

\begin{document}
\def\BLoRa{$\text{B}^2$LoRa\xspace}
\title{\BLoRa: Boosting LoRa Transmission for Satellite-IoT Systems with Blind Coherent Combining}
\vspace{-5mm}
\author{
    Yimin Zhao\textsuperscript{1}\textsuperscript{*},
    Weibo Wang\textsuperscript{1}\textsuperscript{*},
    Xiong Wang\textsuperscript{2}$^{\dagger}$,
    Linghe Kong\textsuperscript{1}$^{\dagger}$,\\
    Jiadi Yu\textsuperscript{1}, 
    Yifei Zhu\textsuperscript{1},
    Shiyuan Li\textsuperscript{1},
    Chong He\textsuperscript{1},
    Guihai Chen\textsuperscript{1}
}
\affiliation{%
    \institution{
        \textsuperscript{1}Shanghai Jiao Tong University 
        \hspace{0.6cm} 
        \textsuperscript{2}University of Science and Technology of China
    }
}

\begin{abstract}
With the rapid growth of Low Earth Orbit (LEO) satellite networks, satellite-IoT systems using the LoRa technique have been increasingly deployed to provide widespread Internet services to low-power and low-cost ground devices. However, the long transmission distance and adverse environments from IoT satellites to ground devices pose a huge challenge to link reliability, as evidenced by the measurement results based on our real-world setup. In this paper, we propose a blind coherent combining design named \BLoRa to boost LoRa transmission performance. The intuition behind \BLoRa is to leverage the repeated broadcasting mechanism inherent in satellite-IoT systems to achieve coherent combining under the low-power and low-cost constraints, where each re-transmission at different times is regarded as the same packet transmitted from different antenna elements within an antenna array. Then, the problem is translated into aligning these packets at a fine granularity despite the time, frequency, and phase offsets between packets in the case of frequent packet loss. To overcome this challenge, we present three designs — joint packet sniffing, frequency shift alignment, and phase drift mitigation to deal with ultra-low SNRs and Doppler shifts featured in satellite-IoT systems, respectively. Finally, experiment results based on our real-world deployments demonstrate the high efficiency of \BLoRa.
\renewcommand{\thefootnote}{\relax}
\footnotetext{\textsuperscript{*} Yimin Zhao and Weibo Wang are the co-primary authors.}
\footnotetext{$^{\dagger}$ Xiong Wang and Linghe Kong are the co-corresponding authors.}
\renewcommand{\thefootnote}{\arabic{footnote}}
\end{abstract}


\fancyhead[LE]{\sffamily\footnotesize ACM MOBICOM '25, November 4--8, 2025, Hong Kong, China}
\fancyhead[RE]{\sffamily\footnotesize Y. Zhao, W. Wang, X. Wang, L. Kong, J. Yu, Y. Zhu, S. Li, C. He, G. Chen}
\fancyhead[LO]{\sffamily\footnotesize \BLoRa: Boosting LoRa Transmission for Satellite-IoT Systems}
\fancyhead[RO]{\sffamily\footnotesize ACM MOBICOM '25, November 4--8, 2025, Hong Kong, China}
\maketitle

\vspace{-3mm}
\section{Introduction}
Recent years have witnessed the rapid growth of LEO satellite networks~\cite{liu2024democratizing, L2D2, mobicom24li, mobicom23li}.  Given the global coverage, LEO satellites have been employed to connect widely distributed IoT devices. Many companies \cite{company_SWARM, company_FOSSA, company_SATELIOT, company_Myriota, company_EchoStar, company_ORBCOM, company_astrocast} have already entered the satellite-IoT market by launching cost-effective nanosatellites~\cite{OEC}. LoRa~\cite{LR-sensitivity}, as one of the most noise-resistant low-power wireless protocols, is widely utilized to facilitate communication between these nanosatellites and IoT devices.

Despite the distinguishable noise-resistant ability, the LoRa link still faces the issue of ultra-low Signal-to-Noise Ratio (SNR) and significant Doppler frequency shift, which stem from the long link distance ranging from 500 to \SI{2500}{\unit{\kilo\meter}} as well as the extreme mobility. To address this issue, Semtech develops LR-FHSS~\cite{LR-FHSS} as a potential replacement for LoRa. However, its pseudo-random frequency-hopping mechanism imposes substantial computational overhead on receivers, limiting its application to uplinks only~\cite{LR-FHSS-gateway} (devices to satellites), while LoRa downlink optimization remains unsolved. Moreover, downlinks in satellite-IoTs are highly frequent and crucial. Unlike the relatively static nature of terrestrial networks, the rapidly changing topologies necessitate that satellites send downlink beacons to inform IoT devices of future passes, enabling timely uplink establishment.

To tackle this issue, operational IoT satellites typically adopt a \textit{repeated broadcasting mechanism} (i.e., re-transmitting at regular intervals)~\cite{Spectrumize} to broadcast the same downlink beacon over a period, thereby achieving a higher reception rate. Nonetheless, our real-world measurements (\S\ref{sec: Motivation}) reveal that it still encounters a severe packet loss rate of up to 90\%, indicating that unreliable downlinks remain a significant problem in satellite-IoT systems.

To deal with this problem, there exist two representative research works --- Spectrumize~\cite{Spectrumize} and XCopy~\cite{XCopy}. Specifically, Spectrumize~\cite{Spectrumize} is designed for ground stations in satellite-IoT systems, aiming at enhancing LoRa packet detection capability by eliminating the pre-calculated Doppler frequency shift. However, it can not be applied to IoT devices, as cold-started IoT devices lack prior information about the dynamic satellite ephemeris to pre-compute the Doppler shift. Additionally, Spectrumize does not address the payload decoding problem in low SNR conditions, which frequently occurs in satellite-IoT systems. XCopy~\cite{XCopy} is designed to improve LoRa link performance by combining re-transmitted packets in terrestrial networks, yet it can not solve the inevitable Doppler shift present in satellite-IoT systems. Under ultra-low SNR conditions, XCopy also fails to detect a sufficient number of packets for effective combining.

In this paper, we propose a \textbf{B}lind coherent combining design to \textbf{B}oost the \textbf{LoRa} link performance for satellite-IoT systems, referred to as \textbf{\BLoRa}. Leveraging the repeated broadcasting mechanism in satellite-IoT systems, \BLoRa achieves coherent combing of re-transmitted packets from satellites to boost the SNR of packets received at IoT devices without requiring prior knowledge of channel states or satellite orbital locations. However, implementing \BLoRa presents challenges in terms of two aspects.

First, the ultra-long link distance up to \SI{2500}{\unit{\kilo\meter}} introduces severe attenuation. Furthermore, satellite-IoT's low-power and low-cost features prohibit conventional SNR enhancement operations, such as increasing transmission power or using large antenna arrays for beamforming. This results in ultra-low SNRs, causing frequent packet loss.

\begin{figure}[t]
    \centering
    \begin{subfigure}{0.23\textwidth}
        \centering
        \includegraphics[width=\textwidth]{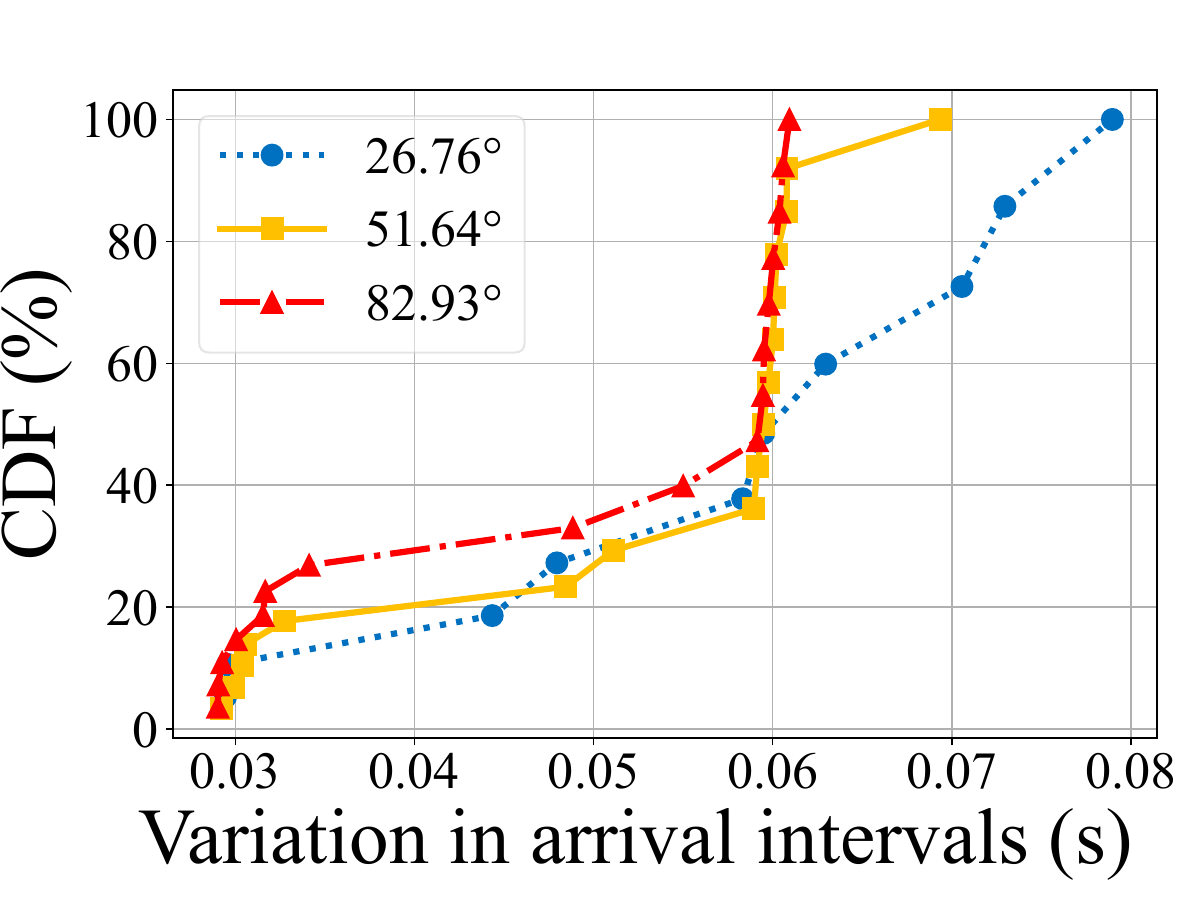}
        \vspace{-10mm}
        \captionsetup{justification=raggedright, singlelinecheck=false}
        \caption{}
        \label{fig: interval}
    \end{subfigure}
    \hfill
    \begin{subfigure}{0.23\textwidth}
        \centering
        \includegraphics[width=\textwidth]{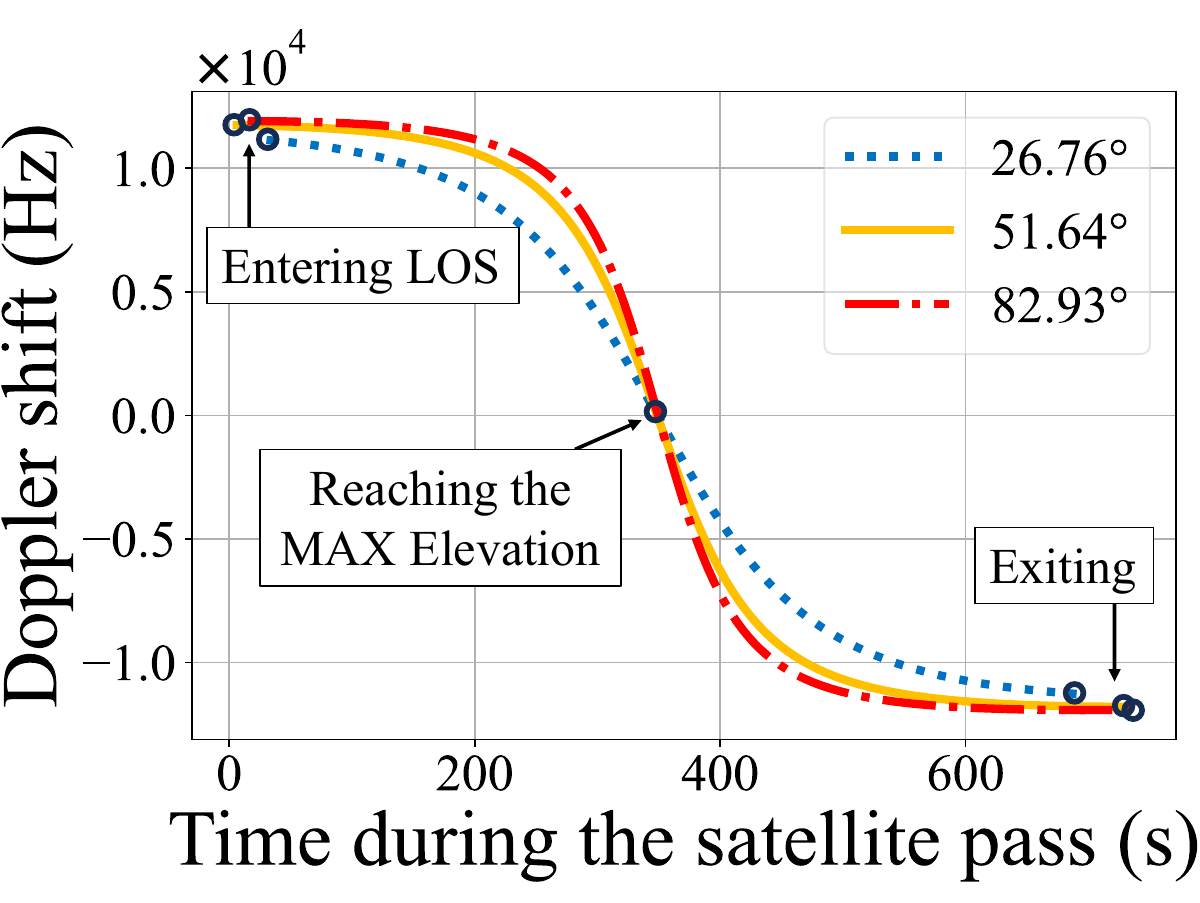}
        \vspace{-10mm}
        \captionsetup{justification=raggedright, singlelinecheck=false}
        \caption{}
        \label{fig: doppler}
    \end{subfigure} 
    \caption{Measurement results on our real-world testbed during three satellite passes at varying MAX elevation angles: 
        (a) The Cumulative Distribution Function (CDF) of the variation in packet arrival intervals between re-transmissions; 
        (b) The changing Doppler shift (when the carrier frequency is 503 MHz).
    }
\end{figure}

Second, due to the swift orbital motion of satellites and hardware imperfections, packets experience inherent misalignment in arrival time, frequency, and phase. To illustrate, measurements based on the real-world testbed reveal two phenomena.
(1) \textit{The packet arrival intervals at ground devices are irregular}, despite the IoT satellite's regular transmission period, caused by varying propagation times and clock inaccuracies, as shown in Figure~\ref{fig: interval}.
(2) The Doppler frequency shift follows a center-symmetric variation pattern, peaking when the satellite enters the line-of-sight (LOS), dropping to zero at maximum elevation, and then symmetrically declining into negative values as it exits LOS, as illustrated in Figure~\ref{fig: doppler}.
\textit{The constantly changing Doppler shift can result in inter-packet frequency shift of up to tens of kHz for narrowband protocols such as LoRa. Meanwhile, the varying Doppler shift even exists within packets}, impacting \BLoRa's performance. Additionally, the inevitable Carrier Frequency Offset (CFO) between satellites and devices exacerbates the frequency shift problem. In satellite-IoT systems where the Spreading Factor (SF) and Bandwidth (BW) are respectively set as 11 and \SI{125}{\unit{\kilo\hertz}} to maintain the LoRa link, a tiny time difference of $8\times$\num{e-6} \unit{\second} or a subtle frequency difference of \SI{61}{\hertz} can incur destructive combining.

Subsequently, we present the design of \BLoRa aimed at overcoming the challenges mentioned above.

To detect packets under ultra-low SNRs, we propose a joint packet sniffing scheme that leverages the repeated broadcast mechanism, allowing for packet detection even when the individual packet is indistinguishable from noise. Specifically, we modify the standard LoRa packet detection scheme by prolonging the preamble detection window to accumulate more energy per packet. Next, according to the repeated broadcast mechanism, we reframe the detection problem as a target recognition problem in Synthetic Aperture Radar (SAR) to achieve joint packet detection. In this way, \BLoRa can detect packets that are deeply buried in noise and obtain their precise arrival time.

To address the frequency shift problem, we design a frequency shift alignment scheme to tackle the misalignment of inter-packet Doppler shift, intra-packet Doppler shift variations, and CFO without requiring orbital data. First, our analysis unveils that the intra-packet Doppler shift variations can be approximated as linear. Subsequently, we perform conjugate multiplication between two re-transmitted packets, which cancels out all symbol components and produces a chirp-like signal caused by the CFO and Doppler shift. We demonstrate that the initial frequency of this chirp-like signal represents the sum of the inter-packet Doppler shift and CFO misalignment, while the slope indicates the variation rate misalignment of the intra-packet Doppler shift. By compensating for this chirp-like signal, we achieve frequency shift alignment between the two packets.

Next, we tackle the phase drift problem between packets, primarily caused by imperfections in the Phase-Locked Loop (PLL) within LoRa hardware. We reveal the two reasons behind the imperfections: the frequency response delay at the frequency hopping points of chirps and the random-phase-startup defect. For the frequency response delay, we demonstrate that it has a minimal impact on coherent combining due to the structural similarity of re-transmitted packets. Concerning the random-phase-startup defect, we propose an initial phase search method to address this problem. By performing \( n \) rounds of active compensation, the initial phase difference between packets can be constrained within \( \left( - \frac{\pi}{n}, \frac{\pi}{n} \right) \), facilitating coherent combining with an acceptable computation overhead.

Up to now, we have shown the process of frequency and phase shift alignments between two packets. Then, we select one detected packet as the anchor and align all other detected packets with this anchor packet. Therefore, \BLoRa achieves blind coherent combining under ultra-low SNRs and Doppler shifts while requiring no prior knowledge of satellite ephemeris and ground device locations.

Finally, extensive experiments are conducted based on our real-world testbed with an in-orbit nanosatellite as well as a Software Defined Radio (SDR)-based testbed. 
Results demonstrate that \BLoRa provides a 9 dB gain for packet detection compared to standard LoRa and improves the packet reception ratio by 190\% over standard LoRa, 80\% over XCopy, and 45\% over Spectrumize. 

The contributions of this paper are summarized below.
\begin{itemize}
   \item We propose a blind coherent combining design named \BLoRa, leveraging the repeated broadcasting mechanism in satellite-IoT systems for link enhancement.
   \item To deal with the ultra-low link budgets, we design a joint packet sniffing algorithm to find out the missing packets deeply buried in the noises. 
   \item To achieve accurate alignment of packets influenced by Doppler shifts and hardware imperfections, we devise a frequency shift alignment scheme and a phase drift mitigation method, ultimately enabling the coherent combination of these packets.
   \item We carry out extensive experiments to verify the high efficiency of \BLoRa based on both a real-world satellite-IoT testbed and an SDR-based testbed.
\end{itemize}

\section{Background and Motivation}
\subsection{Background}
To facilitate a better understanding of \BLoRa, we first introduce the physical layer protocol --- LoRa. Then, we present a high-level description of LoRa-based satellite-IoT systems.

\textbf{LoRa primer.} LoRa~\cite{jkadbear/gr-lora} is a physical (PHY) layer protocol based on Chirp Spread Spectrum (CSS) modulation. All LoRa packets are constructed with \textit{chirps}, where an \textit{up-chirp}/\textit{down-chirp} sweeps the frequency spectrum increasingly/decreasingly. The time-on-air of the chirp, denoted as $\mu$, is determined by the Spreading Factor (SF) and the Bandwidth (BW), following $\mu = \frac{2^{SF}}{BW}$.
A \textit{standard} up-chirp in baseband can be represented as 
\begin{equation}
    c(t, f_0) = Ae^{j2\pi(f_0 + \frac{1}{2}kt)t}, 0 \le t < \mu.
\end{equation}
Here, $A$ denotes the amplitude, $f_0 = -\frac{BW}{2}$ represents the initial frequency, and $k = \frac{BW}{\mu}$ is defined as the frequency sweep rate. A symbol in baseband can be expressed as
\begin{equation}
    c(t, f_{sym}) = Ae^{j2\pi f(t, f_{sym})t}, 0 \le t < \mu,
\end{equation} where $f_{sym}$ is a frequency shift representing the symbol, and $f(t, f_{sym})$ is the encoded frequency function defined as
\begin{equation}
    f(t, f_{sym}) =
    \begin{cases}
        f_{sym} + \frac{1}{2}kt,      & 0 < t \le t_{sym};\\ 
        f_{sym} - BW + \frac{1}{2}kt, & t_{sym} \le t <  \mu. 
    \end{cases}
    \label{eq: fchirp}
\end{equation} 
\noindent The \textit{frequency hopping} happens in $t_{sym}$ which is defined as
\begin{equation}
    t_{sym} = \frac{\frac{BW}{2} - f_{sym}}{k}.
\end{equation}
\begin{figure}[t]
  	\centering
  	\includegraphics[width=8cm]{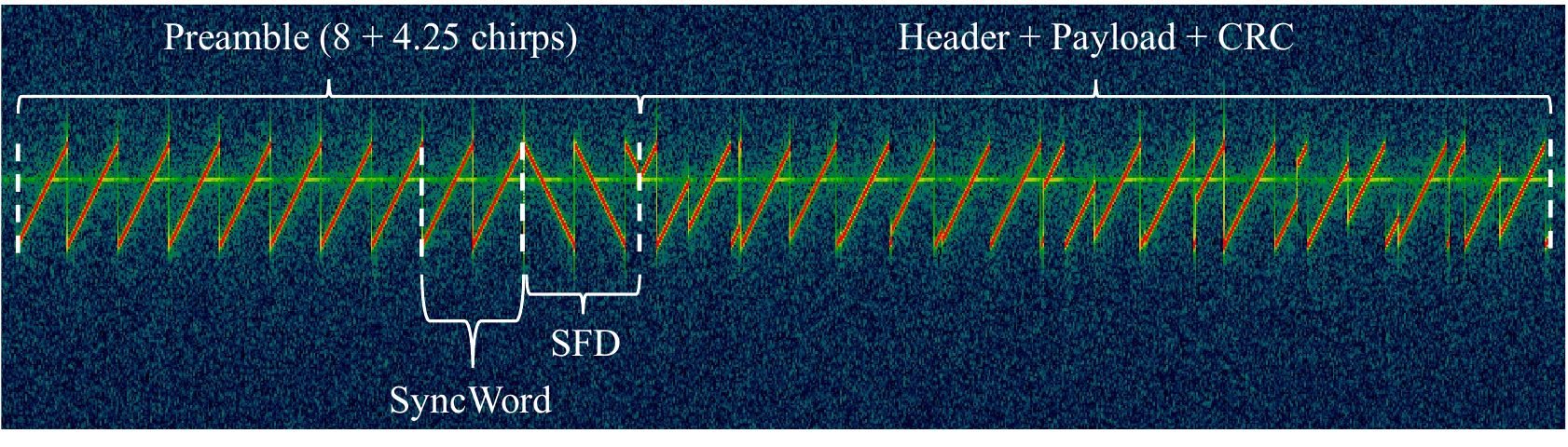}
	\caption{The spectrogram of a LoRa packet.}
    \label{fig: lora_packet}
\end{figure}
Symbols are demodulated by \textit{dechirping}, which involves multiplying the received chirps with a standard down-chirp, followed by a Fast Fourier Transform (FFT) to retrieve the encoded frequency $f_{sym}$. A LoRa packet is structured with a preamble, an optional header, a payload, and an optional CRC. The preamble typically consists of standard up-chirps, a SyncWord, and a Start Frame Delimiter (SFD). Figure~\ref{fig: lora_packet} illustrates an example of a LoRa packet.

\textbf{LoRa-based satellite-IoT system.}
\label{sec: Background}
\begin{figure}[t]
  	\centering
  	\includegraphics[width=8cm]{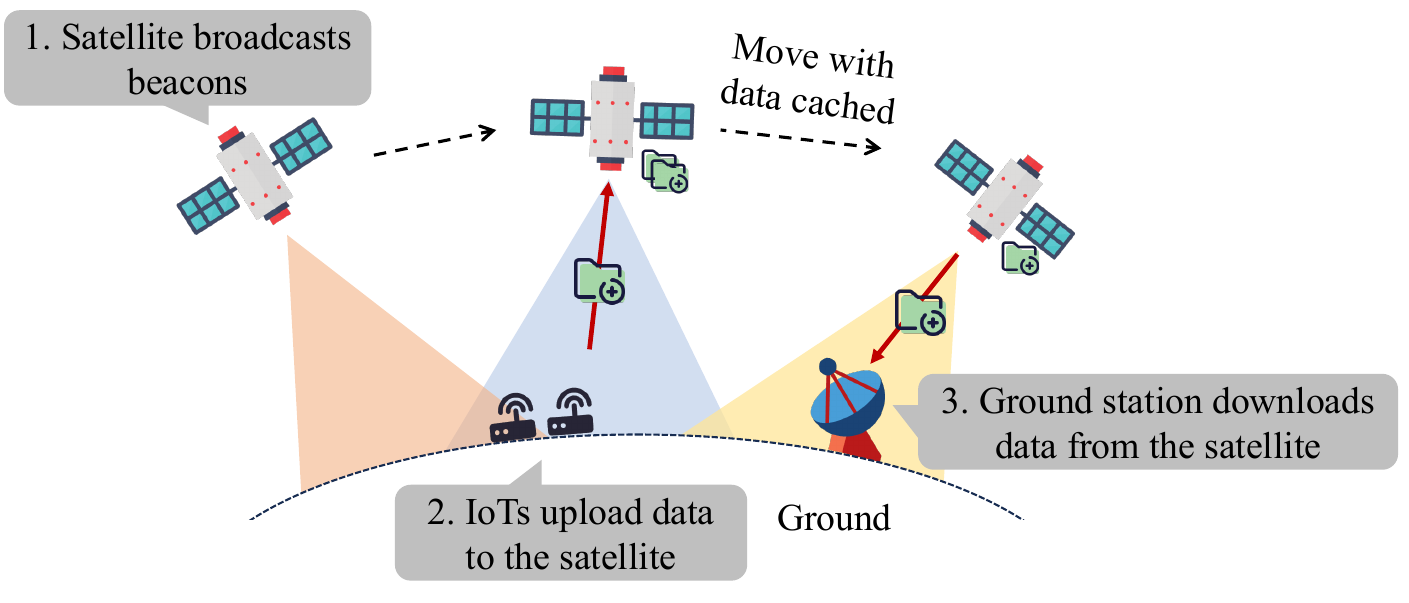}
	\caption{The overview of satellite-IoT systems.}
    \label{fig: sIoT_overview}
\end{figure}
Typically, a satellite-IoT system comprises LEO satellites, IoT devices, and ground stations, as shown in Figure~\ref{fig: sIoT_overview}. LEO satellites act as IoT gateways, while the IoT devices deployed on the ground are responsible for sensing information. In the system's operational workflow, the first step involves the satellite broadcasting beacons to the ground. Most of these beacons contain Two-Line Element (TLE)~\cite{TLE} information, enabling IoT devices that receive them to predict satellite orbits. This prediction allows devices to power down their radio transceivers before the next satellite pass to conserve energy, and then perform data uploads during the subsequent pass. The satellite caches the data uploaded by the IoT devices and downloads it upon passing over a ground station, where the data is then distributed to the tenants/users of the IoT devices.

The satellite-IoT system exhibits two unique characteristics. (1) Compared to terrestrial IoT systems, \textit{the downlink in a satellite-IoT system is more frequent and critical.} This is because satellite gateways move at high speeds, providing each IoT device only a brief window of access. Consequently, satellites must continuously broadcast beacons to convey TLE information, allowing IoT devices worldwide to predict the satellites’ positions and seize the short-lived access opportunities. In contrast, ground gateways can often offer prolonged device access and do not require frequent downlink transmissions. (2) Compared to LEO mega-constellations like Starlink, \textit{a satellite-IoT system can operate with a relatively small number of satellites} (e.g., even a single satellite). This cost-effectiveness arises from the generally low time sensitivity of IoT data retrieval. A single satellite is capable of providing global coverage, serving as a gateway for IoT devices worldwide, and relaying data back to the ground within a day. For IoT users involved in applications such as environmental monitoring, this delay is typically acceptable.

\subsection{Motivation}
\label{sec: Motivation}
\begin{figure}[t]
  \centering
  \begin{subfigure}{0.13\textwidth}
    \centering
    \includegraphics[width=\textwidth]{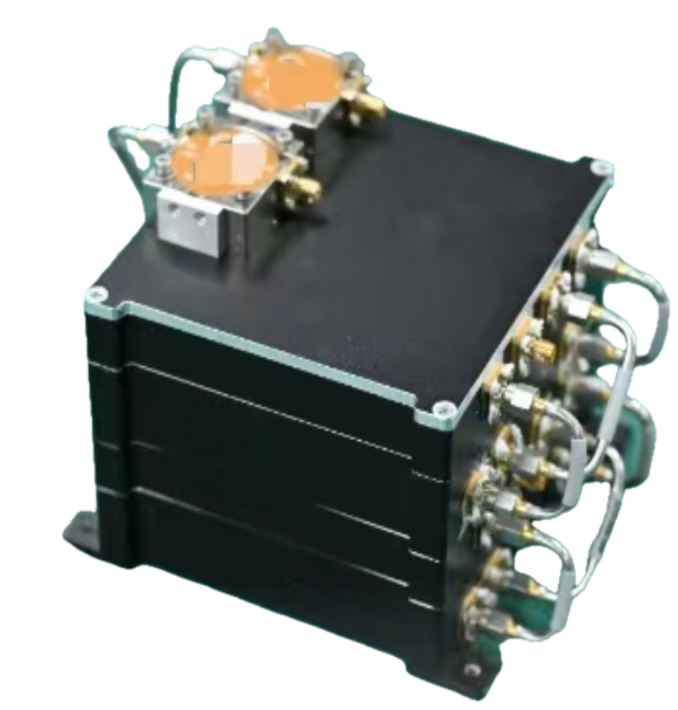}
    \vspace{-3mm}
    \caption{}
    \label{fig: satellite}
  \end{subfigure}
  \hfill
  \begin{subfigure}{0.14\textwidth}
    \centering
    \includegraphics[width=\textwidth]{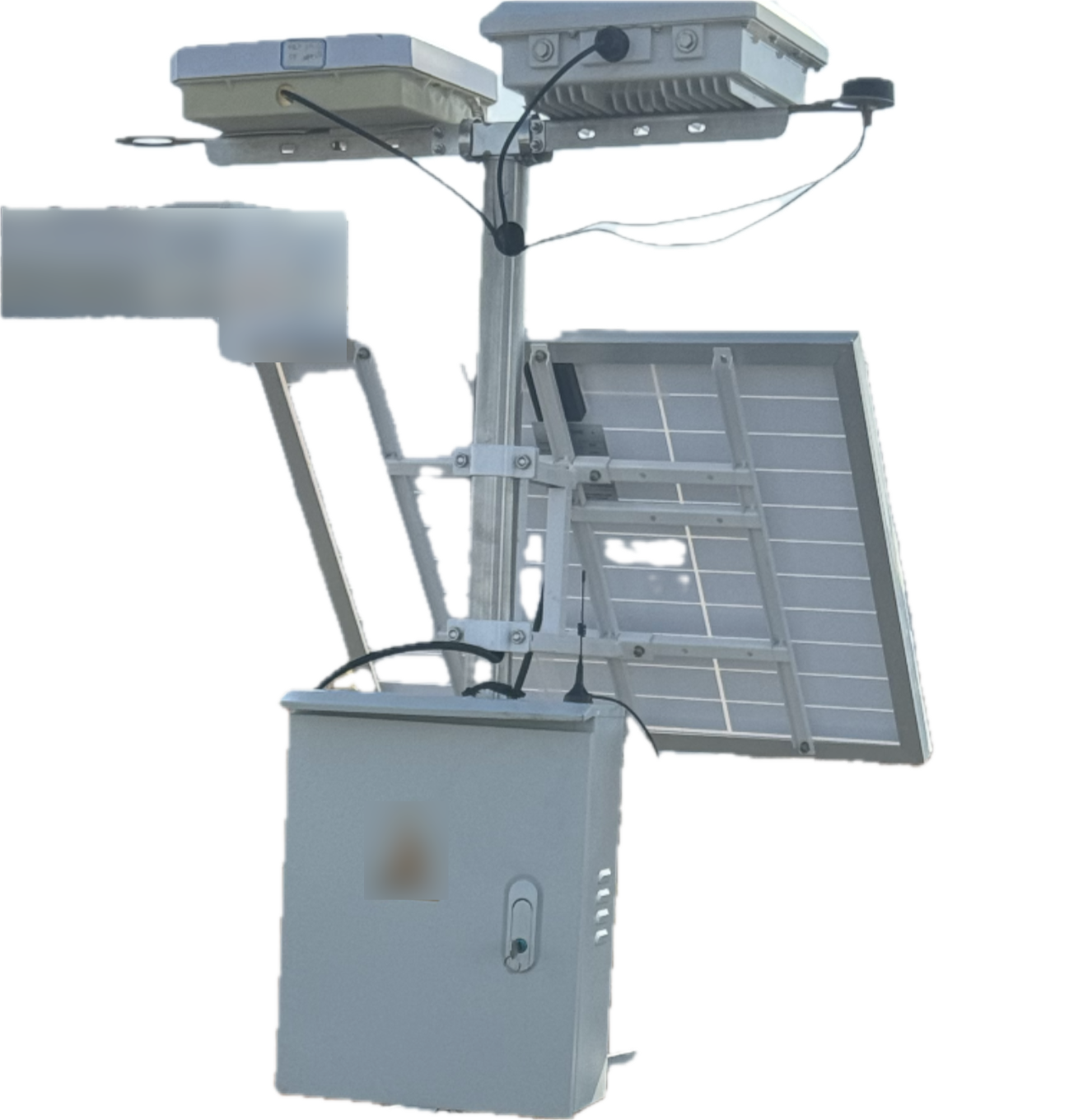}
    \vspace{-3mm}
    \caption{}
    \label{fig: device}
  \end{subfigure}
  \begin{subfigure}{0.19\textwidth}
    \centering
    \includegraphics[width=\textwidth]{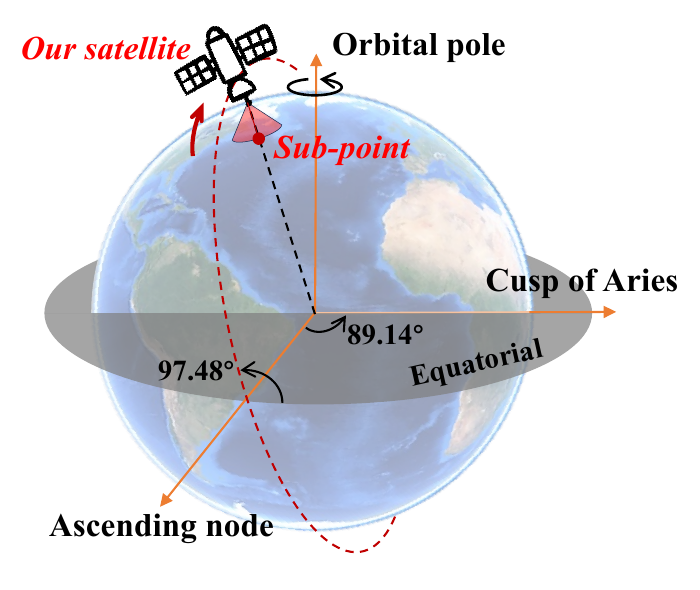}
    \vspace{-3mm}
    \caption{}
    \label{fig: orbit}
  \end{subfigure}
  \begin{subfigure}{0.225\textwidth}
    \centering
    \includegraphics[width=\textwidth]{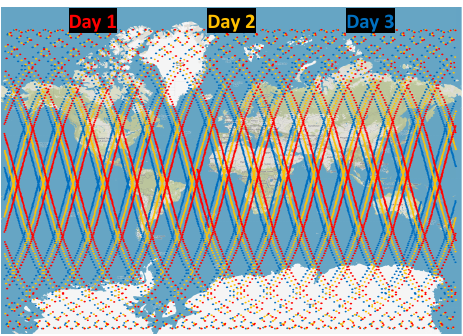}
    \vspace{-4mm}
    \caption{}
    \label{fig: coverage}
  \end{subfigure}
  \begin{subfigure}{0.225\textwidth}
    \centering
    \includegraphics[width=\textwidth]{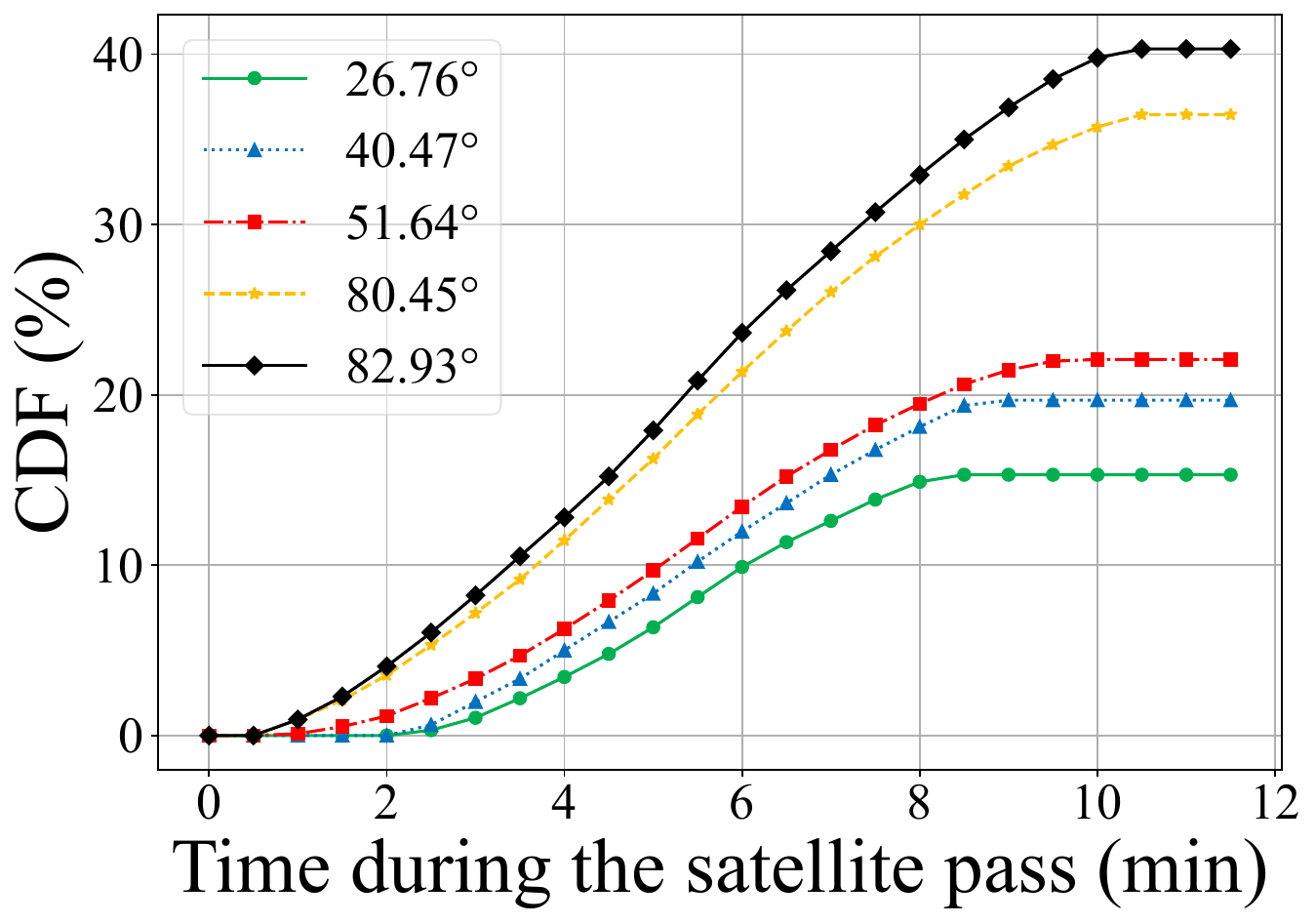}
    \vspace{-4mm}
    \caption{}
    \label{fig: CDF}
  \end{subfigure}
  \begin{subfigure}{8cm}
    \resizebox{\linewidth}{!}{
    \begin{tabular}{r|c|c|c|c}
    \hline
     & 
     \makecell[c]{Reykjavik\\(N\ang{64;08.48})}&
     \makecell[c]{Shanghai\\ (N\ang{31;13.43})}&
     \makecell[c]{Singapore\\(N\ang{1;17.0})}& 
     \makecell[c]{Melbourne\\(S\ang{37;48.49})}\\ 
    \hline
    $80\degree < i \leq 90\degree$ & 0.30/d & 0.14/d & 0.12/d & 0.14/d \\
    \hline
    $50\degree < i \leq 80\degree$ & 1.18/d & 0.61/d & 0.47/d & 0.65/d \\
    \hline
    $20\degree < i \leq 50\degree$ & 2.63/d & 1.23/d & 1.02/d & 1.35/d \\
    \hline
    $0\degree < i \leq 20\degree$  & 6.77/d & 2.62/d & 2.24/d & 2.86/d \\
    \hline
    \end{tabular}
    }
    \caption{}
    \label{fig: passes}
  \end{subfigure}
  \caption{The real-world satellite-IoT testbed:
    (a) Our satellite's LoRa transmitter (Semtech SX1276);
    (b) Our IoT devices with LoRa receiver (Semtech SX1276);
    (c) Our satellite's orbit;
    (d) Our satellite's sub-point trajectory in three days;
    (e) The CDF of an IoT device of the testbed detecting at least one LoRa packet transmitted by our satellite during passes of varying MAX elevation angles;
    (f) The average number of times the satellite passes over four different latitude locations each day, based on data over 90 days, categorized by the MAX elevation angle (denoted as $i$) of each pass.
  }
  \label{fig: setup}
\end{figure}
In LoRa-based satellite-IoT systems, cost-effective and low-power LEO nanosatellites are essential. However, their transmission power is constrained by the small size of solar panels and electromagnetic compatibility constraints. Additionally, beamforming with large-scale antenna arrays is not feasible due to cost considerations. Given the considerable distance between LEO satellites and ground devices, ensuring reliable LoRa connectivity becomes a critical issue.

To explore this, we build a real-world satellite-IoT testbed with an LEO nanosatellite operating at approximately \SI{530}{\unit{\kilo\meter}} altitude and ground IoT devices. Both the satellites and the devices are equipped with commercial LoRa transceivers, Semtech SX1276~\cite{SX1276}, as shown in Figure~\ref{fig: satellite} and \ref{fig: device}. Figure~\ref{fig: orbit} shows our satellite’s actual orbital plane. The satellite broadcasts a LoRa packet (a TLE~\cite{TLE} beacon) to the ground every 30 seconds, in line with the typical broadcast interval~\cite{Spectrumize}. We focus on evaluating the reception performance of TLE beacons. Since these beacons can only be received by ground IoT devices within the satellite's coverage area, it is vital to explore the satellite’s orbital passes first.

Unlike geostationary (GEO) satellites that hold a fixed position relative to Earth, LEO satellites orbit much faster due to lower altitudes. The speed difference between Earth's rotation and the satellite's motion creates a non-linear, non-periodic sub-point (the location directly beneath the satellite, as shown in Figure~\ref{fig: orbit}) trajectory. The sub-point trajectory of our satellite over 3 days in 2024 is recorded and visualized in Figure~\ref{fig: coverage}. We can observe that the satellite flyes over the entire globe daily but with sub-point trajectories changing.

Consequently, the satellite's elevation angle, defined as the angle between the LOS from the ground to the satellite and the horizontal plane, varies significantly with each pass. Lower elevation angles extend the transmission range and degrade the link quality, making it essential to quantify their distribution across diverse passes. We statistically analyze elevation angles during our satellite passing four cities at different latitudes. Over a continuous 90-day period in 2024, we document all passes, as shown in Figure~\ref{fig: passes}. It is observed that during most passes, the elevation angle remains under $20\degree$ regardless of location. This finding is broadly applicable, as our satellite's orbital design reflects the typical characteristics of LEO satellites used for IoT communication, which should ensure global coverage~\cite{LEO-inclination}.

Next, we investigate the actual LoRa link performance based on the elevation angle distribution. We measure the number of packets detected by the IoT device during our satellite passes at varying elevation angles. It's observed that during passes with maximum elevation angles exceeding $80\degree$, the CDF for detecting at least one packet remains below 40\%, as reported in Figure~\ref{fig: CDF}. These high-elevation passes occur on average 0.14 times daily, while most passes have angles below $20\degree$, leading to a reception ratio under 20\%. Therefore, the link performance from satellites to devices is unreliable.

Motivated by the above observations, it is necessary to enhance the received signal on the ground. Given the cold-start mechanism of IoT devices, which lack prior knowledge of satellite trajectories, Channel State Information (CSI) remains unavailable.
Consequently, conventional methods like phased-array beamforming cannot be applied due to the absence of CSI or limitations on power and cost.
Therefore, we propose a blind coherent combining design that utilizes re-transmissions from a single antenna to achieve an accumulation gain akin to the beamforming of a phased array.

The main challenges include:
\begin{itemize}
    \item Ultra-low SNR: The significant link distance of up to \SI{2500}{\unit{\kilo\meter}} results in substantial link loss, rendering most packets difficult to detect, and letting alone obtaining precise arrival times of packets.
    \item Inter/intra-packet frequency shift: Varying Doppler frequency shift occurs due to the changing radial speed from satellites to devices during transmission, leading to substantial shifts even within single packets, exacerbated by Carrier Frequency Offset (CFO); 
    \item Inter-packet phase drift: Hardware imperfections and Sampling Time Offset (STO) introduce phase drift among packets, diminishing coherence during combining and reducing gain.
\end{itemize}
\section{System Design}
\begin{figure}[t]
    \centering
    \includegraphics[width=0.47\textwidth]{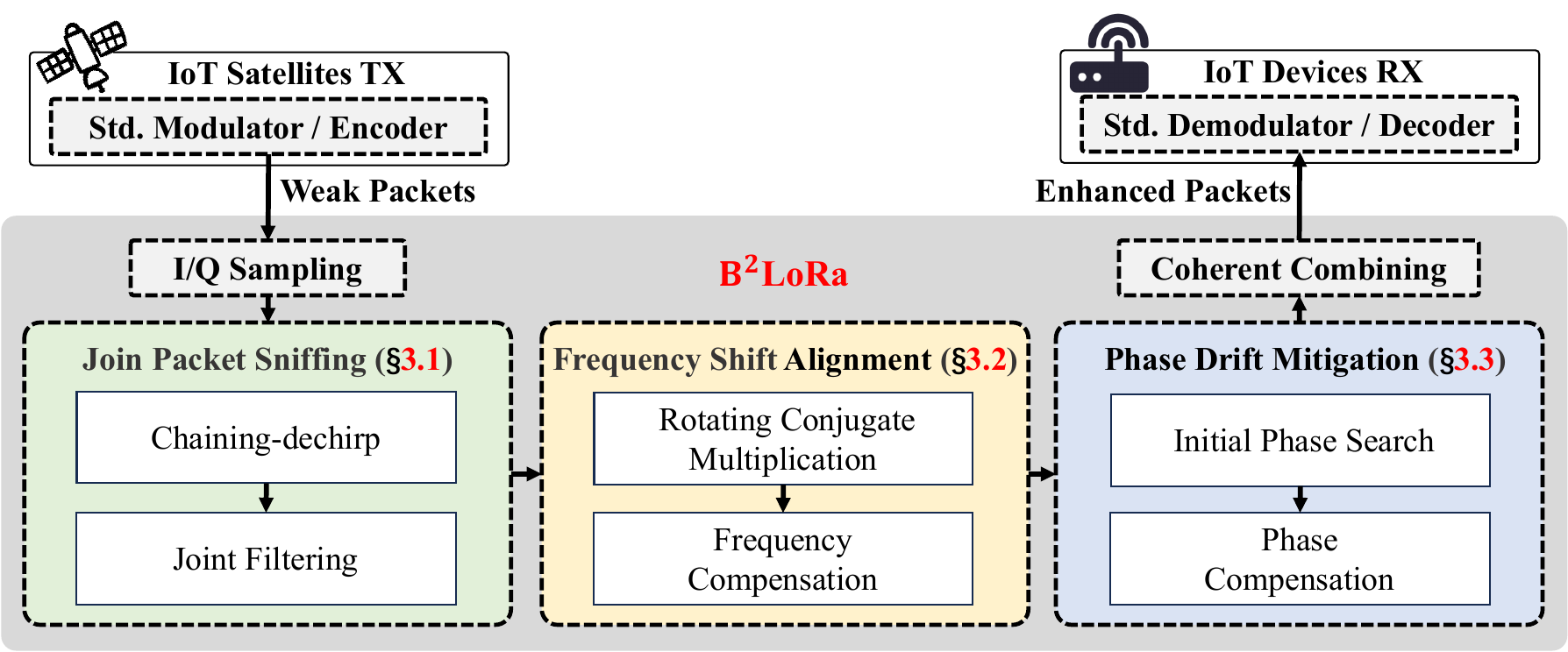}
    \vspace{-2mm}
    \caption{The framework of \BLoRa.}
    \label{fig: framework}
    \vspace{-2mm}
\end{figure}
In challenging satellite-IoT communication conditions, we present a \textbf{B}lind coherent combining design to \textbf{B}oost SNRs, \textbf{\BLoRa}, to enhance \textbf{LoRa} link performance. Figure~\ref{fig: framework} illustrates the framework of \BLoRa. Operating as an overlay layer for LoRa-based satellite-IoT systems, \BLoRa collects, processes raw satellite signals, and delivers enhanced LoRa packets to IoT devices. It consists of three main components:

\textbf{(1) Joint packet sniffing.} \BLoRa begins with utilizing the \textit{chaining-dechirp} technique to aggregate more energy compared to the dechirping in standard LoRa. Then, by leveraging the inter-packet periodicity resulting from the satellite's regular re-transmissions, \BLoRa reformulates the packet detection problem as a target recognition problem in Synthetic Aperture Radar (SAR). Finally, weak packets with their precise arrival times can be jointly detected through the \textit{joint filtering algorithm}.

\textbf{(2) Frequency shift alignment.} Next, \BLoRa aligns packets in the frequency domain. We first demonstrate theoretically and experimentally that, for LoRa packets transmitted by LEO satellites, the Doppler shift within each packet can be approximated as linearly varying. As a result, \BLoRa applies a \textit{rotating conjugate multiplication} method to determine the frequency difference between packets as a linear component, accommodating both inter/intra-packet Doppler shift and CFO. Through frequency compensation, accurate frequency alignment between packets can be achieved.

\textbf{(3) Phase drift mitigation.} Lastly, \BLoRa mitigates the phase drift of re-transmitted packets. We conduct an empirical study to uncover the underlying causes of the phase drift and analyze its distribution and severity within and across packets. As a result, \BLoRa employs the \textit{initial phase search} method to mitigate this issue. Ultimately, coherent combinations of packets are achieved, which boosts SNRs.

\subsection{Joint Packet Sniffing}
\label{sec: joint_packet_sniffing}
Standard LoRa suffers from high packet loss rates in ultra-low SNRs. Since \BLoRa boosts SNRs through the blind coherent combining of re-transmitted packets, maximizing the number of detected packets is vital. To this end, \BLoRa first employs the chaining-dechirp technique to increase accumulated energy in preamble detection. Then, \BLoRa introduces a joint filtering algorithm inspired by SAR to further enhance the packet detection capability.

\textbf{Chaining-dechirp technique.}
\begin{figure}
    \begin{subfigure}[t]{0.5\textwidth}
        \centering
        \includegraphics[width=8.5cm]{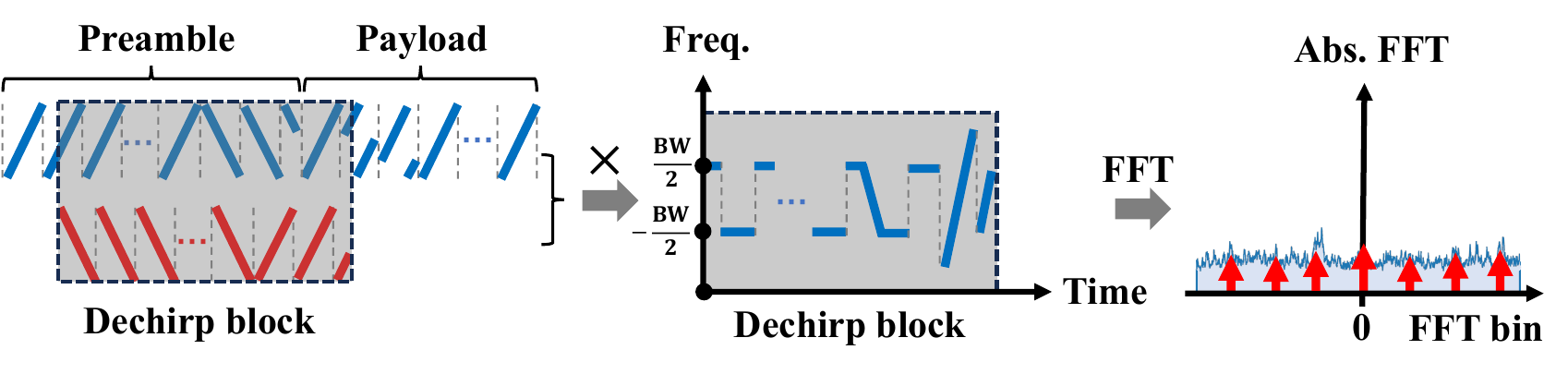}
        \caption{A frequency-varying output.}
        \label{fig: chaining_misalign}
    \end{subfigure}
    \hfill
    \begin{subfigure}[t]{0.5\textwidth}
    \centering
        \includegraphics[width=8.5cm]{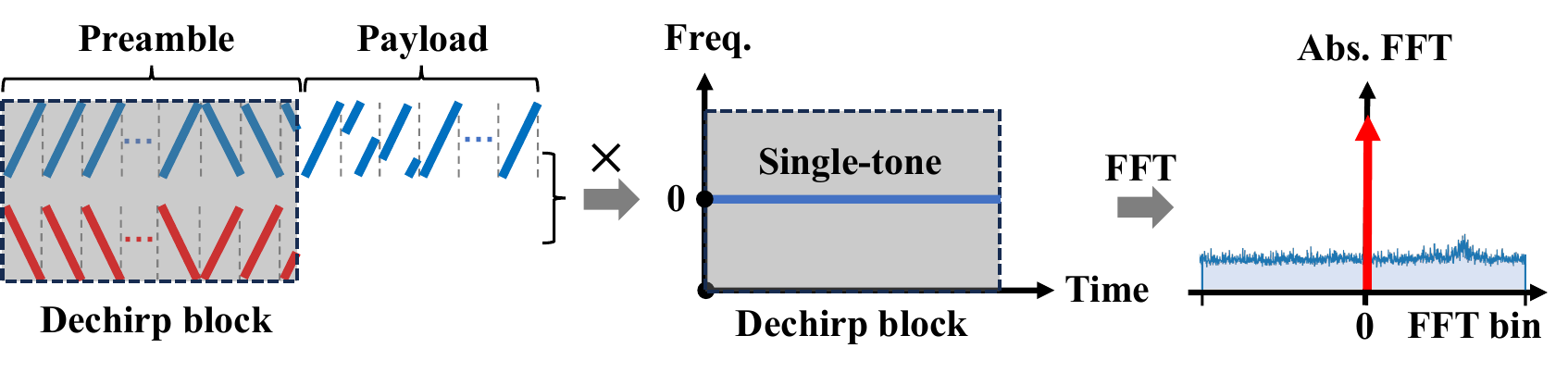}
        \caption{A single-tone output.}
        \label{fig: chaining}
    \end{subfigure}
    \caption{Illustration of two types of outputs when employing sliding chaining-dechirp on received signals.}
    \label{fig: chaining_all}
\end{figure}
The standard LoRa protocol identifies a packet by detecting the periodicity of consecutive up-chirps within its preamble. This detection occurs through a dechirping process that utilizes a sliding window, generating periodic FFT peaks that correspond to the periodic up-chirps in the preamble as it slides through the packet, indicating the packet’s presence. However, in satellite-IoT systems, the link budget is considerably lower, often causing these FFT peaks to fall below the noise floor. To address this challenge, we propose the chaining-dechirp technique instead of the standard dechirping. By leveraging the fact that the preamble structure is prior-known information in current satellite-IoT systems, this technique uses the conjugate transformation of the entire preamble as a dechirp block.

Upon receiving signals, \BLoRa employs sliding chaining-dechirp, which involves sliding the dechirp block over sampling points and performing multiplication. When the dechirp block and the packet preamble are not aligned, as shown in Figure~\ref{fig: chaining_misalign}, the resulting signal has multiple frequency components, yielding no significant FFT peaks. When perfectly aligned, as illustrated in Figure~\ref{fig: chaining}, the energy from all chirps focuses on the carrier frequency. This can generate a significant FFT peak, which signifies the presence of a packet. Compared to dechirping, which only gathers energy from a single chirp, chaining-dechirp can combat lower SNRs.

\textbf{Joint filtering algorithm.}
\begin{figure}[t]
    \begin{subfigure}{0.23\textwidth}
        \centering
        \includegraphics[width=\textwidth]{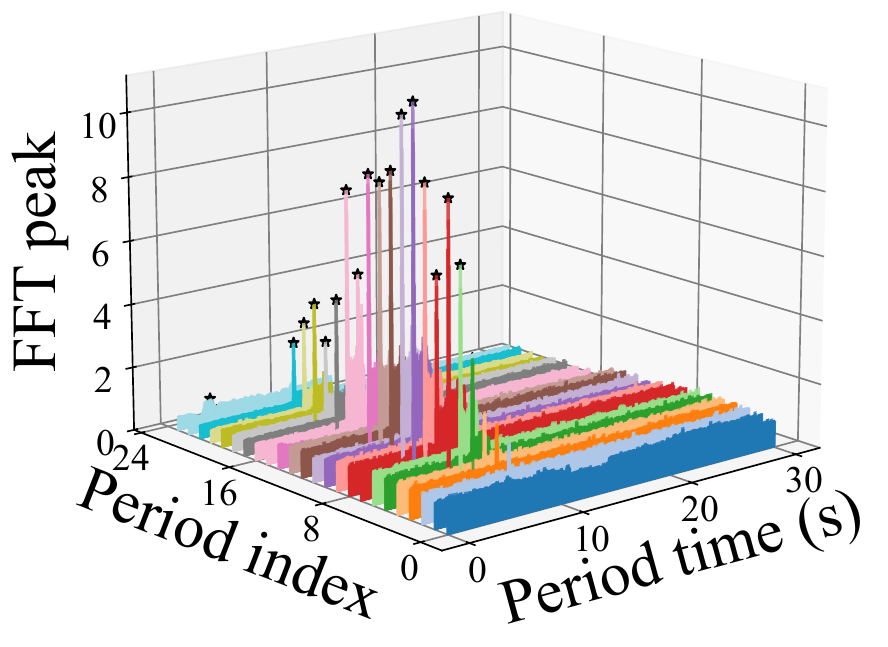}
        \vspace{-10mm}
        \captionsetup{justification=raggedright, singlelinecheck=false}
        \caption{}
        \label{fig: fft_3D}
    \end{subfigure}
    \begin{subfigure}{0.23\textwidth}
        \centering
        \includegraphics[width=\textwidth]{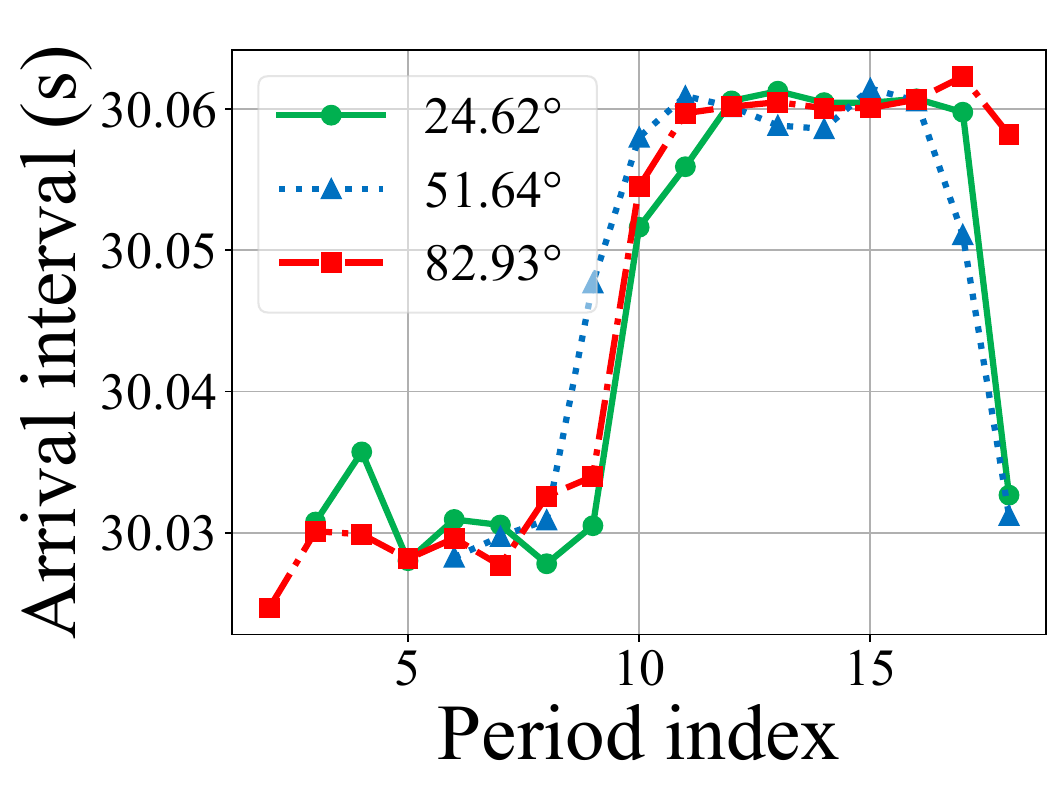}
        \vspace{-10mm}
        \captionsetup{justification=raggedright, singlelinecheck=false}
        \caption{}
        \label{fig: position_diff}
    \end{subfigure}
    \caption{(a) The heatmap acquired through a sliding chaining-dechirp process; (b) The arrival intervals of periodically re-transmitted packets.}
\end{figure}
Although chaining-dechirp can accumulate a higher FFT peak for each packet, relying solely on this single peak for packet detection is not optimal. As the LEO satellite approaches or exits the LOS, its distance from the receiver can increase by approximately $5\times$, leading to significantly lower SNRs for packets during these times. This may cause even the FFT peak from chaining-dechirp to drop below the noise level. However, an intriguing feature is that the repeated broadcasting mechanism of IoT satellites closely resembles that of \textit{Synthetic Aperture Radar} (SAR) satellites, which also periodically transmit pulse signals to the ground during their operation and utilize this periodicity for echo detection and analysis. Inspired by SAR, we propose a joint filtering algorithm on top of sliding chaining-dechirp to further enhance packet detection capability.

Specifically, while performing sliding chaining-dechirp on the received signal, an FFT spectrum is generated at each sliding step, from which \BLoRa records the maximum value (i.e., the FFT peak). This yields a time series of FFT peaks representing the power spectrum's maximum values throughout the satellite's pass. Given that satellites broadcast at a fixed period ($\tau$), this time series is then segmented into several periods based on $\tau$. This process can be defined as:
\begin{equation}
    \begin{aligned}
        &i = \left\lfloor{\frac{t}{\tau}}\right\rfloor, \quad t_{p} = t \bmod {\tau}, \quad 0 \le t \le T,
        \vspace{-2mm}
    \end{aligned}
\end{equation}
where $T$ is the duration of the satellite pass, $i$ represents the index of a period, $t_p$ denotes the time within a period.

A heatmap can be generated with $i$ and $t_p$ as axes, using the FFT peak at $i\cdot \tau + t_p$ as the value. Figure~\ref{fig: fft_3D} shows an example based on our testbed, where the broadcast period $\tau$ is \SI{30}{\second}. This heatmap is essentially a SAR image, showing periodic \textit{peaks of FFT peaks} for re-transmitted packets, marked by $\star$ in Figure~\ref{fig: fft_3D}. At this point, the packet detection problem transforms into linear target detection within the SAR image. Generally, during the satellite pass, the FFT peaks of these packets initially rise, then fall, with initial and final phases suffering from weak energy indistinguishable from noise. Additionally, significant channel uncertainty at long distances may cause some packets to deviate from this trend and be obscured by noise. The joint filtering algorithm addresses this by exploiting packet periodicity, \textit{using stronger adjacent packets to infer weaker, noise-obscured ones.} 

It begins by denoising the SAR image matrix $\bm{H}$, where $\bm{H}[m][n]$ represents the FFT peak at $i = m$ and $t_p = n$. To eliminate the noise points without compromising LoRa signals, points whose values are close to the average value are discarded. This is based on the fact that most energy comes from background noise, while LoRa packets occupy minimal time and energy. Thus, the mean of each row or column of $\bm{H}$ approximates background noise energy. Taking the $m$-th row in $\bm{H}$ for example, \BLoRa discards values in $\bm{H}[m]$ below $k\times \overline{\bm{H}[m]}$, where $\overline{\bm{H}[m]}$ is the average for the $m$-th row and $k$ is the filtering threshold. Empirically, a threshold of $k = 1.5$ effectively balances noise reduction and signal preservation.

After reducing noise, the joint filtering algorithm searches for periodic LoRa packet emergence within the denoised $\bm{H}$. However, periodicity is imperfect due to varying propagation times as the satellite moves, as shown in Figure~\ref{fig: position_diff}, which illustrates packet arrival intervals across three satellite passes. Thus, detecting periodicity must account for temporal relaxation. \BLoRa achieves this through linear regression, attempting to find a line in $\bm{H}$ (i.e., the SAR image) that minimizes the loss function defined as
\begin{align}
    \vspace{-2mm}
    \L = \sum_{m,n} d[m][n] \cdot \bm{H}[m][n]^2,
    \vspace{-2mm}
\end{align} 
where $(m,n)$ represents the points where the line intersects, and $d[m][n]$ represents the horizontal distance from each point to the fit line. By leveraging prior knowledge of rough periodicity, indicating the fit line should be nearly vertical in $\bm{H}$, we incorporate horizontal distance $d[m][n]$ into the loss metric. This approach significantly accelerates convergence time and reduces computational cost. Through linear regression, \BLoRa obtains the coarse arrival times of all packets (including those obscured by noise), which correspond to the intersection point of the fitted line with the SAR image.

The acquired coarse arrival times still experience subtle inter-packet time offsets. Specifically, our orbital calculation and empirical data reveal that (1) relativistic effects introduce a negligible time deviation between the satellite and ground device of about \num{e-10}\unit{\second} per second; (2) the variation in transmission distance caused by satellite motion results in an  \num{e-3}\unit{\second} difference in propagation time between re-transmissions; and (3) discrepancies between satellite/device's circuit delays may result in deviations of  \num{e-6} to \num{e-3}\unit{\second} per second. \BLoRa employs a well-studied method, the inter-packet sliding conjugate multiplication~\cite{XCopy}, to mitigate inter-packet time offset to less than a single sample ($< 4\times$ \num{e-6}\unit{\second}). Since linear regression already constrains this offset within a few samples, \BLoRa's computational complexity of time offset alignment significantly decreases, making the deployment for IoT devices more feasible.

\subsection{Frequency Shift Alignment}
\label{sec: FO alignment}
\begin{figure}[t]
    \begin{subfigure}{0.20\textwidth}
        \centering
        \includegraphics[width=\textwidth]{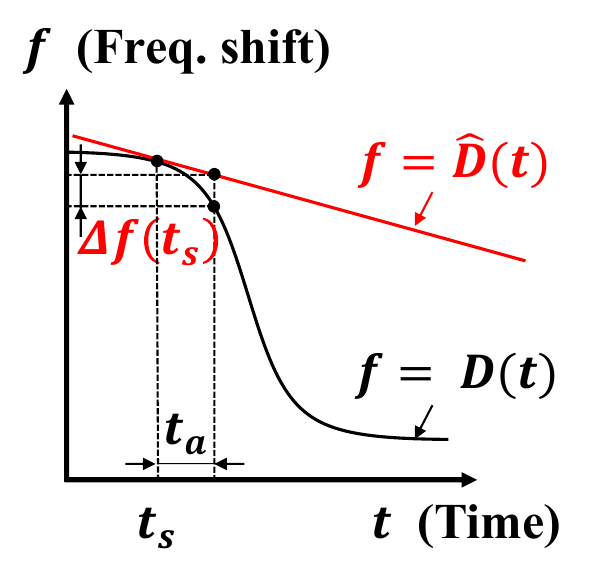}
        \vspace{-10mm}
        \captionsetup{justification=raggedright, singlelinecheck=false}
        \caption{}
        \label{fig: doppler_error}
    \end{subfigure}
    \begin{subfigure}{0.26\textwidth}
        \centering
        \includegraphics[width=\textwidth]{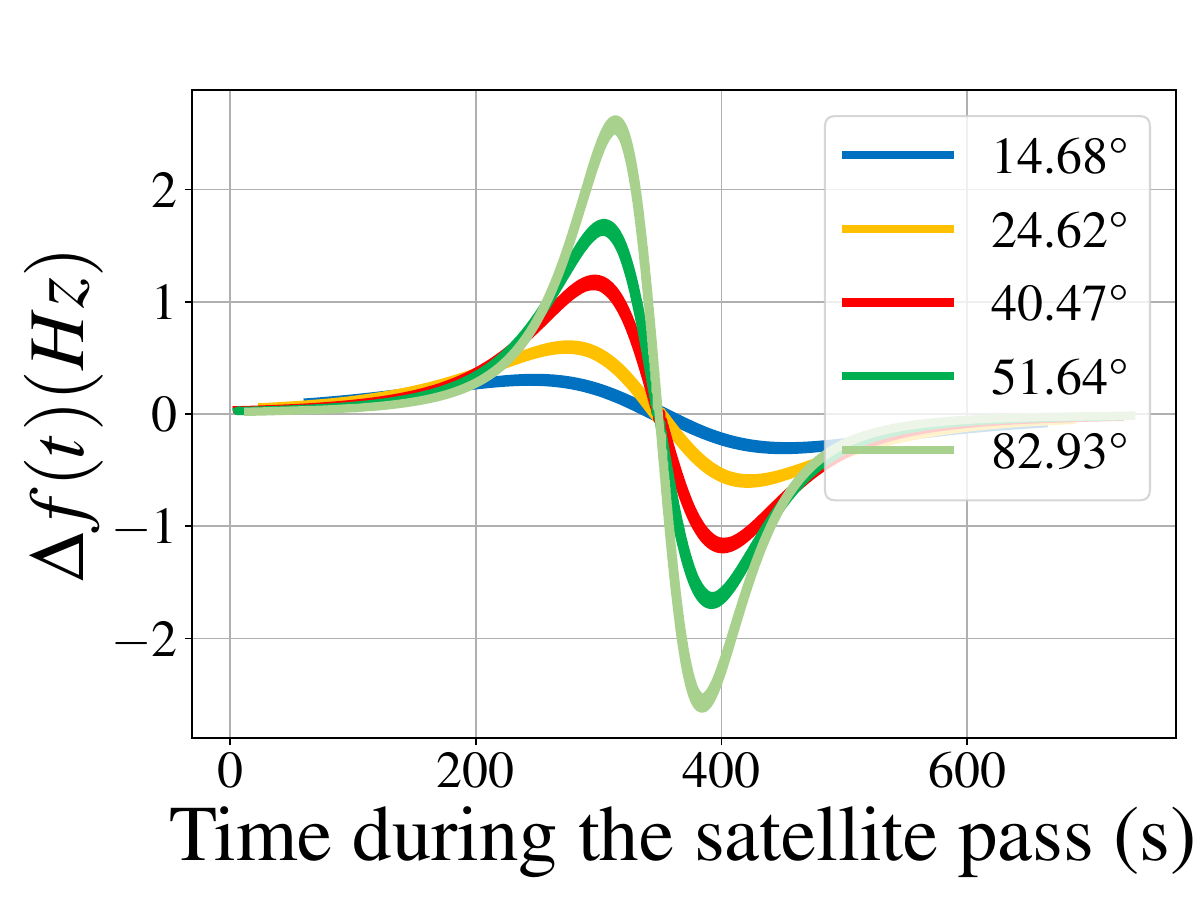}
        \vspace{-10mm}
        \captionsetup{justification=raggedright, singlelinecheck=false}
        \caption{}
        \label{fig: delta_f}
    \end{subfigure}
    \vspace{-3mm}
    \begin{subfigure}[t]{8cm}
        \resizebox{\linewidth}{!}{
            \begin{tabular}{r|r|r|r|r}
            \hline
                       & \textbf{1000 MHz} & \textbf{800 MHz} & \textbf{500 MHz} & \textbf{400 MHz}\\ 
            \hline
            \textbf{200 km} & 39.01~\unit{\hertz} & 31.21~\unit{\hertz} & 19.50~\unit{\hertz} & 15.60~\unit{\hertz} \\
            \hline
            \textbf{500 km} & 5.60~\unit{\hertz} & 4.48~\unit{\hertz} & 2.80~\unit{\hertz} & 2.24~\unit{\hertz} \\
            \hline
            \textbf{1000 km} & 1.15~\unit{\hertz} & 0.92~\unit{\hertz} & 0.58~\unit{\hertz} & 0.46~\unit{\hertz} \\
            \hline
            \textbf{2000 km} & 0.20~\unit{\hertz} & 0.16~\unit{\hertz} & 0.10~\unit{\hertz} & 0.08~\unit{\hertz} \\
            \hline
            \end{tabular}
        }
        \vspace{-2mm}
        \caption{}
        \label{fig: error_all_case}
    \end{subfigure}
    \caption{
        (a) The error (i.e., $\Delta f(t_s)$) introduced by the linear approximation of intra-packet DFS; 
        (b) The change of $\Delta f(t)$ when $t = t_s$ during passes of varying MAX elevation angles;
        (c) The various MAX values of $\Delta f(t_s)$ throughout 90$\degree$ satellite passes with different satellite altitudes (rows) and carrier frequencies (columns).
    }
    \vspace{-2mm}
\end{figure}

\begin{figure}[t]
    \centering
  	\includegraphics[width=8.7cm]{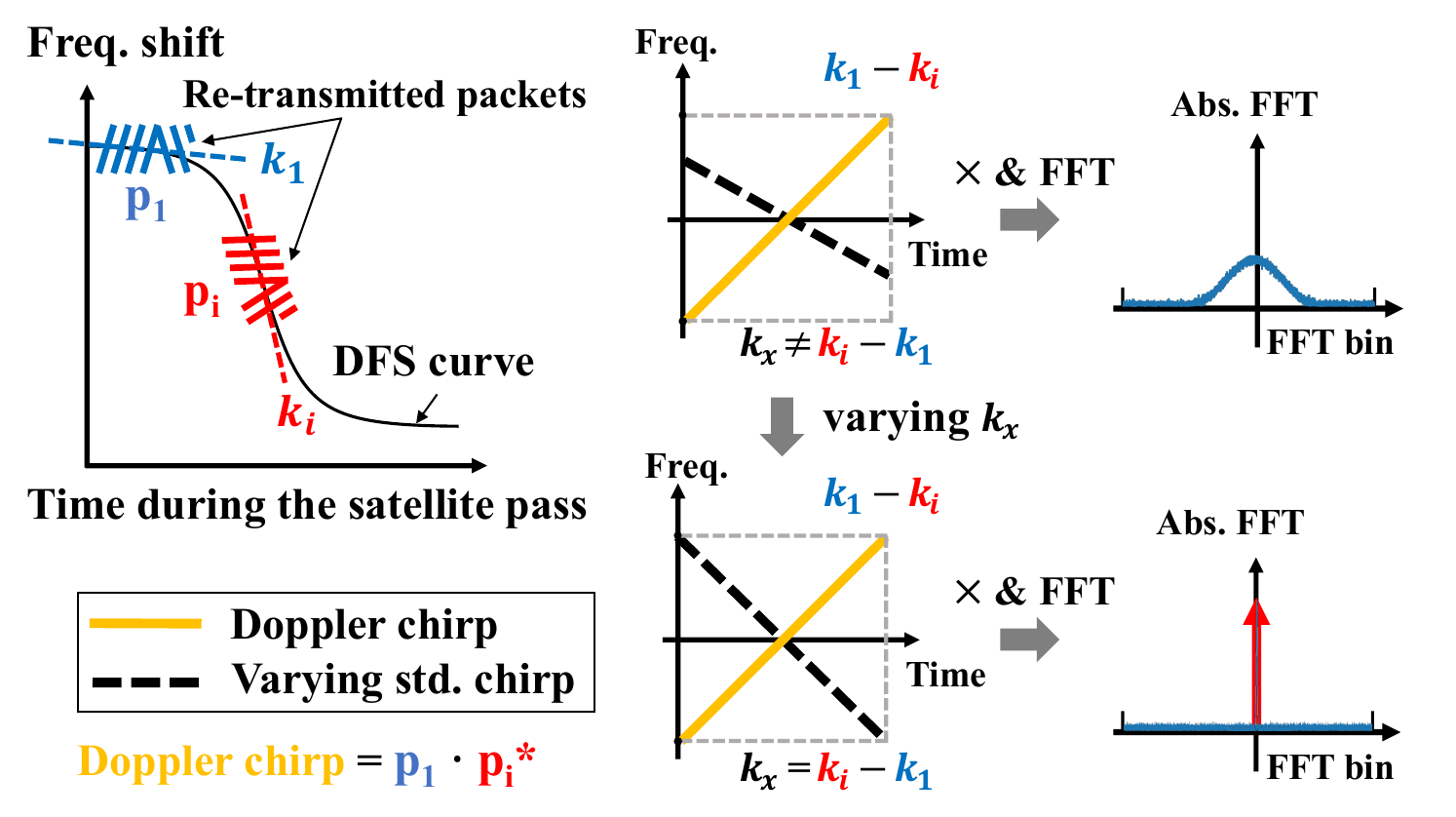}
	\caption{Illustration of the construction of the Doppler chirp and finding its slope through dechirping it with a slope (denoted as $k_x$)-varying standard chirp.}
    \label{fig: doppler_cancellation}
\end{figure}

After detecting packets and obtaining their precise arrival times, \BLoRa next handles the time-varying Doppler frequency shift (DFS) between and within packets, as long as CFO.
Due to the lack of satellite orbital information, conventional Doppler compensation schemes based on astrodynamics calculations are not applicable to IoT devices.

Consequently, \BLoRa leverages the consistent content of re-transmitted packets to create a new scheme, a rotating conjugate multiplication algorithm, that directly derives the frequency difference between packets via inter-packet operations without requiring prior satellite information. 
This algorithm is based on our important finding, as shown in Proposition~\ref{prop}. We first prove it and then elaborate on the design of the rotating conjugate multiplication algorithm.
\vspace{-2mm}
\begin{proposition}
    The intra-packet DFS can be approximated as a linear change with time without inducing decoding errors for commercial LoRa-based LEO satellites.
    \label{prop}
    \vspace{-2mm}
\end{proposition}
\textbf{Proof.}
We begin by modeling the errors caused by this approximation and then prove that these errors do not impact LoRa decoding.
As shown in Figure~\ref{fig: doppler_error}, let $t$ denote the time during the satellite pass, $f$ represent the frequency shift, and $D(t)$ be the observed DFS at time $t$. 
For a packet starting at $t_s$ with a time-on-air of $t_a$, the maximum intra-packet DFS is $D(t_s + t_a) - D(t_s)$.
\BLoRa approximates the \( D(t) \) curve between \( t=t_s \) and \( t=t_s+t_a \) as the tangent line at \( t=t_s \) (i.e., the linear approximation of the DFS). Let this line be \( f = \hat{D}(t) \); since it passes through \( (t_s, D(t_s)) \), it can be expressed as \( f = \hat{D}(t) = D^{'}(t_s)t + D(t_s) - D^{'}(t_s)t_s \), where \( D^{'}(t_s) \) is the first derivative of \( D(t) \) at \( t_s \).

Then, the difference between the approximated intra-packet DFS and the ground-truth intra-packet DFS for packet starting at time $t_s$, denoted as $\Delta f(t_s)$, can be expressed as
\begin{equation}
\begin{aligned}
    \Delta f(t_s) &= \hat{D}(t_s+t_a) - D(t_s + t_a)\\
                &= D^{'}(t_s)(t_s+t_a) + D(t_s) - D^{'}(t_s)t_s - D(t_s + t_a).
\label{eq: delta_f}
\end{aligned}
\end{equation}

The greater the maximum elevation angle during the satellite pass, the more severe the DFS variation. Under a fixed elevation angle, DFS solely depends on the satellite’s altitude and carrier frequency. Accordingly, we conduct orbital calculations for LEO satellites within an altitude range of 200 km to 2000 km and for common legitimate LoRa frequency bands (400 MHz to 1000 MHz)~\cite{LR-regional-para}. The aim was to verify whether $|\Delta f(t_s)|$ remains sufficiently small to avoid symbol errors during LoRa demodulation. Figure~\ref{fig: error_all_case} presents the maximum values of $|\Delta f(t_s)|$ experienced during passes at a maximum elevation angle of $90\degree$, varying across different altitudes (rows) and different carrier frequencies (columns). The results indicate that (1) this maximum value increases as altitude decreases and carrier frequency increases, and (2) this maximum value remains below the error threshold $BW/2^{SF}$ ($\approx$61 Hz when SF=11 and BW=125kHz). Consequently, within a wide range of configurations for LoRa-based LEO satellites, the linear approximation of DFS does not induce errors in symbol demodulation. Additionally, Figure~\ref{fig: delta_f} displays the $|\Delta f(t_s)|$ as a function of pass time for our real-world satellite (altitude of 530 km, carrier frequency of 503 MHz) under various passes. It's observed that the maximum value is significantly lower than the error threshold.

\textbf{Rotating conjugate multiplication.}
Without loss of generality, we model the received signal of the \textit{i}-th detected packet encoded by $f(t)$ as
\begin{equation}
     p_i(t)=A_i e^{j\{2\pi[f(t) + D(t) + f_{cfo[i]}]t + \phi_i(t)\}} + n(t),
     \label{eq: linear_F}
\end{equation} 
where $A_i$ is the amplitude affected by the communication channel, $D(t)$ represents the DFS, $f_{cfo[i]}$ is CFO, $\phi_i(t)$ represents the packet-unique phase drift caused by factors such as hardware imperfection and STO, and $n(t)$ denotes noise. The re-transmitted packets have identical content, leading to the same pattern of $f(t)$ within them.

Based on Proposition~\ref{prop}, the DFS across a packet can be approximated as linearly changing, and the DFS at any time $t$ within the $i$-th packet can be expressed as
\begin{equation}
     D(t) = D(t_{s[i]}) + D^{'}(t_{s[i]}) \cdot (t - t_{s[i]}), t_{s[i]} \le t \le t_{s[i]} + t_a,
     \label{eq: linear_DFS}
\end{equation}
where $t_{s[i]}$ is the starting time of the $i$-th packet and $t_a$ represents its time-on-air.
Then, by combining Equation~\ref{eq: linear_F} with Equation~\ref{eq: linear_DFS}, the \textit{conjugate multiplication} result of the first and the $i$-th packets can be obtained as
\begin{equation}
    \label{eq: FO_conjugate}
    p_1(t)\cdot p^{*}_i(t)=A_1  A_{i} e^{j\{2\pi [\Delta D(t) + \Delta f_{cfo}]t + \phi_1 - \phi_{i}\}} + \Delta n(t),
\end{equation}
where $\Delta f_{cfo}$ represents the CFO difference between the two packets and $\Delta n(t)$ denotes the additional terms caused by noise from the two packets. 
Furthermore, $\Delta D(t)$ is
\begin{align}
    \Delta D(t) &= \underbrace{D(t_{s[1]}) - D(t_{s[i]})}_{\text{Inter-packet DFS}} \notag \\
                &\qquad + \underbrace{\left[D^{'}(t_{s[1]}) - D^{'}(t_{s[i]})\right]}_{\text{Difference of intra-packet DFS changing rate}} \cdot t, \notag \\
                 &\quad t_s \le t \le t_s + t_a. \label{eq: Delta_D}
\end{align}

Notably, from Equation~\ref{eq: Delta_D}, it can be observed that the frequency difference between two packets, $\Delta D(t) + \Delta f_{cfo}$, demonstrates a linearly changing feature with time --- in other words, it can be classified as a chirp signal. Therefore, we refer to $p_1(t)\cdot p^{*}_i(t)$ as the \textit{Doppler chirp}. The frequency difference is then found by differentiating this Doppler chirp, leading to two main tasks: (1) Calculate the chirp's frequency-time slope for the intra-packet DFS change rate, and (2) Determine the frequency intercept for the inter-packet DFS and CFO difference.
To find the slope and intercept, \BLoRa \textit{rotates} a standard chirp with varying slopes to dechirp the Doppler chirp. A single-tone signal with the highest FFT peak occurs only when this chirp matches the Doppler chirp. Consequently, the Doppler chirp's slope is the negative of the standard chirp's slope, and its intercept corresponds to the FFT bin of the single-tone signal. This method, illustrated in Figure~\ref{fig: doppler_cancellation}, helps to obtain the time-varying frequency difference between the $i$-th and the first packet, facilitating frequency alignment.

In this way, \BLoRa can combine all packets into one that exhibits time-frequency characteristics consistent with the first packet involved in the combination. As the SNR significantly enhances through combining, \BLoRa calculates the combined packet's frequency offset and linear frequency shift relative to the carrier frequency by dechirping on each up-chirp. This process captures the residual frequency offset/shift, thereby achieving full frequency compensation.

\subsection{Phase Drift Mitigation}
\label{sec: phase}
\begin{figure}
    \centering
    \begin{subfigure}[t]{0.46\textwidth}
        \centering
        \includegraphics[width=\textwidth]{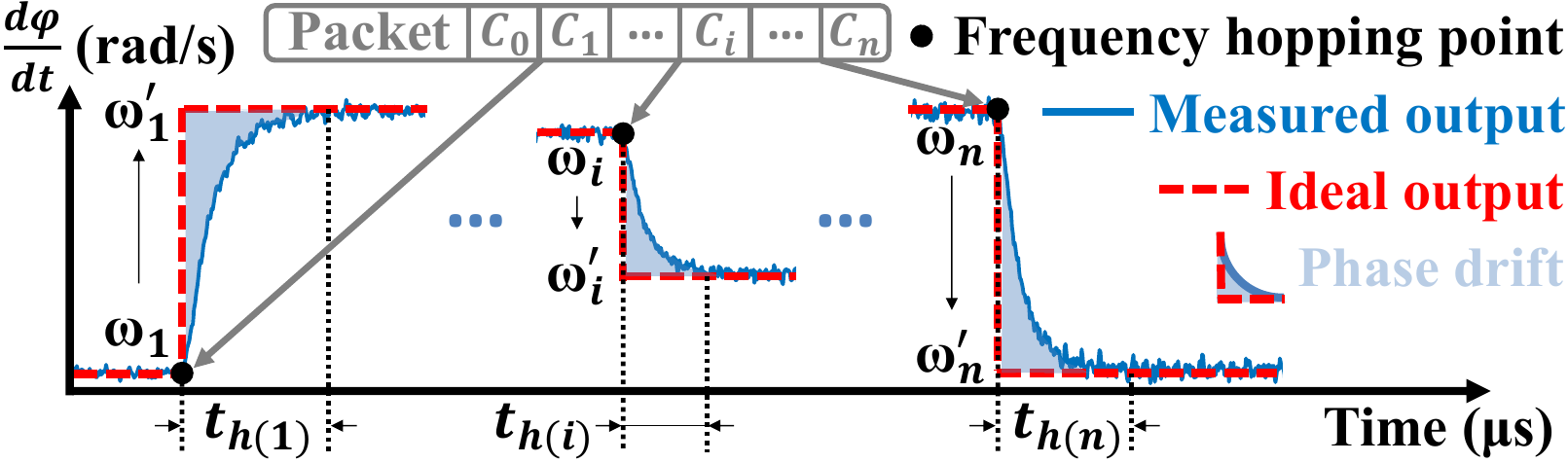}
        \vspace{-5mm}
        \caption{}
        \label{fig: Phase_Drift}
    \end{subfigure}
    \begin{subfigure}[t]{0.46\textwidth}
        \resizebox{\linewidth}{!}{
            \begin{tabular}{c|c|c|c|c}
            \hline
            \makecell[c]{Frequency hopping \\ value (kHz)}&
            \makecell[c]{Phase drift \\ MV (rad)}&
            \makecell[c]{Phase drift \\ RSD}&
            \makecell[c]{$t_h$ \\ MV (\unit{\micro\second})}&
            \makecell[c]{$t_h$ \\ RSD}\\
            \hline
            +62.5 to -62.5 & 5.226 & 3.03\% & 15.625 & 3.13\% \\
            \hline
            +53.6 to -13.6 & 2.129 & 3.10\% & 12.001 & 3.23\% \\
            \hline
            -25.8 to +59.7 & -3.278 & 3.12\% & 13.672 & 3.16\%\\ 
            \hline
            \end{tabular}
        }
        \vspace{-2mm}
        \caption{}
        \label{fig: MV_RSD}
    \end{subfigure}
    \vspace{-3mm}
    \caption{
        (a) Intra-packet phase drift stems from PLL delays at frequency hopping points at the start of chirps (e.g., $C_1$, $C_i$, $C_n$);    
        (b) The Mean Value (MV) and Relative Standard Deviation (RSD) of phase drift and delay time $t_h$ observed during different frequency hops.}
    \vspace{-3mm}
\end{figure}
After time and frequency alignment, resolving phase drift remains essential. Since packet combining essentially involves the vector addition of each sample in the complex plane, maintaining a consistent phase ensures maximum resultant magnitude. Phase drift is categorized into two types: intra-packet phase drift and inter-packet phase drift. We first demonstrate that the intra-packet phase drift has minimal impact on the coherence of combinations and then outline a scheme to mitigate the inter-packet phase drift in \BLoRa.

\textbf{Intra-packet phase drift.}
Phase drift within a LoRa packet primarily occurs at frequency hopping points, both between and within chirps. Our analysis of LoRa hardware~\cite{LoRaWAN, SX1276} and experimental data reveals that phase drift stems from the \textit{frequency response delay} at these points. For commercial LoRa transceivers (e.g., Semtech SX1276/7/8/9), the conversion of the LoRa digital signal into the output analog RF signal relies on adjusting the fractional divider ratio in the feedback loop of the Phase-Locked Loop (PLL). When frequency hopping of a digital signal occurs, the PLL requires time to adjust the divider and loop filter to re-lock the frequency of the output analog signal. As shown in Figure~\ref{fig: Phase_Drift}, during the re-locking interval (e.g., $t_{h(1)}, t_{h(i)}, t_{h(n)}$), the PLL cannot fully follow the ideal waveform, leading to a discrepancy between the actual and ideal changing of frequencies. This delay in frequency response results in phase accumulation deviations, manifesting as phase drift. This phenomenon also exists in other RF instruments like TI mmStudio~\cite{Ti_mmStudio}. Essentially, the magnitude of phase drift is mainly determined by the PLL's response delay, which depends on the frequency-hopping values. Consequently, for any frequency hop of the same value, the same PLL is expected to produce a similar magnitude of phase drift.

To confirm this empirically, we utilize a data capture card to collect nearly noise-free $100\times$ re-transmitted packets from an SX1276 transceiver. Then, we analyze the frequency hops with the same values across these packets. It's observed that the Relative Standard Deviation (RSD) of phase drift and delay time for identical frequency hops is around 3\%, as listed in Figure~\ref{fig: MV_RSD}, indicating that the phase drift variation pattern among re-transmitted packets is highly consistent. Furthermore, we measure the impact of phase drift on the time-on-air of different chirps. The results demonstrate that the RSD of the deviation from the theoretical time-on-air ($2^{SF}/BW$) is around 0.4\%, suggesting that the influence of phase drift is almost confined within each chirp and does not accumulate across symbols. Therefore, \BLoRa only requires the identification of the initial phase difference between two packets to achieve phase alignment throughout them, which is further discussed below.

\textbf{Inter-packet phase drift.}
An initial phase difference (i.e., the inter-packet phase drift) exists between the re-transmitted packets. In addition to the well-studied Sampling Time Offset (STO)~\cite{bernier2020complexity, XCopy}, which contributes to this inter-packet phase drift, we identify the primary factor as \textit{the random-phase-startup defect} of the PLL. Specifically, when a packet is re-transmitted, the PLL restarts to lock the output signal from the Voltage-Controlled Oscillator (VCO) to the input reference signal. This locking process takes time, and the starting phase of the reference signal from the Local Oscillator (LO) also varies. These two aspects introduce randomness in the phase at the start of LoRa packet transmission, leading to a random inter-packet phase drift.

Given its random nature, \BLoRa actively seeks to compensate for the inter-packet phase drift using \(n\) different compensation schemes. In the \(i\)-th scheme, \BLoRa compensates one of the combined packets with the \(i\)-th element from the set \( \left\{ \frac{2\pi}{n}, \frac{4\pi}{n}, \dots, \frac{2n\pi}{n} \right\} \) as the phase. After compensation, \BLoRa combines two packets, followed by dechirping and FFT. The scheme yielding the highest FFT peak is retained because it aligns most closely with the true phase drift. Through this method, \BLoRa can limit the range of inter-packet phase drift to \( \left( - \frac{\pi}{n}, \frac{\pi}{n} \right) \), significantly increasing the combining gain. \BLoRa can perform finer-grained phase compensation as $n$ increases, but this comes at the cost of higher computation overhead. We reveal that $n = 4$ strikes a good balance. In summary, to mitigate inter-packet phase drift, an $n$-element initial phase search is employed, followed by phase compensation. By selecting different search coefficients \( n \) based on the acceptable computational overhead, the optimal combining gain can be achieved accordingly.

\section{Evaluation}
\label{sec: evaluation}
In this section, we evaluate the performance of \BLoRa based on the real-world testbed and the SDR-based testbed.
\vspace{-3mm}
\subsection{Methodology}
\textbf{Prototype implementation.}
The prototype of \BLoRa is implemented on an Ettus X310 USRP~\cite{Ettus}, serving as the RF interface, while a laptop functions as the baseband processor. Since \BLoRa operates as middleware in existing LoRa-based satellite-IoT systems, it takes raw satellite-transmitted signals as input, enhances them, and feeds them into the standard LoRa demodulator/decoder. The demodulator/decoder is implemented with the open-sourced project gr-lora~\cite{jkadbear/gr-lora}.

\textbf{Broadcasting mechanism.}  
During the development phase of satellite-IoT systems, operators may integrate metadata such as timestamps and counters in LoRa packets for bug tracing (i.e., enabling the debug mode), which leads to inconsistent payloads among re-transmissions and hinders coherence of combining. In the evaluation, the debug mode is turned off for deploying \BLoRa in our real-world testbed.

\textbf{Real-world testbed.}
Experiments are conducted on a real-world satellite-IoT testbed, where a satellite equipped with Semtech SX1276 LoRa transceivers orbits at an altitude of \SI{530}{\kilo\meter} (Figure~\ref{fig: satellite}, \ref{fig: orbit}). The satellite features an omnidirectional antenna with a gain of \SI{0}{\si{dBi}} and operates at a transmission power of \SI{5}{\watt}. The LoRa PHY configuration is SF=11, BW=\SI{125}{\kilo\hertz}, at a carrier frequency of \SI{503}{\mega\hertz}. The satellite follows the conventional \textit{repeated broadcasting mechanism}, re-transmitting the same LoRa packet to the ground every 30 seconds. We use an X310 USRP with the planar antenna of an IoT device (Figure~\ref{fig: device}) to capture the raw I/Q data of these signals on the ground. Different intensities of Gaussian noise are added to create a dataset for evaluating \BLoRa. Under ultra-low SNRs, the momentary SNR exhibits significant fluctuations due to overwhelming random noise. To accurately describe the dataset, we use the Received Signal-to-Gaussian Ratio (RGR), defined as the ratio of the power of the received raw signals (including noise) to the energy of the added Gaussian noise, to quantify the relative signal strength of the packets precisely. The raw I/Q data is obtained over 10 months (August 2023 to June 2024), totaling about 720 minutes from 60 orbital passes.

\textbf{SDR-based testbed.}
Due to legal restrictions, wireless parameters can not be arbitrarily adjusted in operational satellites. To evaluate \BLoRa with different PHY parameters, an SDR-based testbed is established. On this testbed, LoRa signals of varying PHY parameters are generated by an SX1276 transceiver on the ground and collected by an X310 USRP. Based on the orbital simulation of our satellite, we calculate the variation of the Free Space Path Loss (FSPL), the Doppler shift, and packet arrival time during various passes. Then, generated LoRa signals are altered according to these factors to ensure they possess SNR, arrival time, and frequency shift characteristics similar to the data on the real-world testbed (phase drift already exists due to similar hardware imperfections). The orbital simulation relies on the SGP4 propagator~\cite{SGP4, skyfield}, with the satellite's latest ephemeris downloaded from CelesTrack~\cite{CelesTrack}.

\textbf{Metrics.}
We evaluate \BLoRa using three metrics:
(1) \textit{Symbol Error Rate (SER)} indicates the ratio of erroneous symbols to the total number of symbols within packets and functions as a metric to measure packet decoding performance;
(2) \textit{Packet Reception Ratio (PRR)} represents the ratio of detected LoRa packets to the total number of packets and is employed as a metric to measure packet detection capability;
(3) \textit{RGR threshold} of detection/decoding represents the minimum RGR that a packet can be detected/decoded and is employed as a metric to measure noise immunity.

\textbf{Comparisons.}
\BLoRa is evaluated against three existing designs:
(1) \textit{LoRa}~\cite{LoRaWAN} represents the standard LoRa protocol, featuring the conventional packet detection logic;
(2) \textit{XCopy}~\cite{XCopy} improves SNRs through coherent combinations of re-transmitted packets in terrestrial links, which is state-of-the-art for terrestrial LoRa networks;
(3) \textit{Spectrumize}~\cite{Spectrumize} enhances the preamble detection capability of ground stations for LoRa satellites, which is the latest packet detection enhancement design for LoRa-based satellite-IoT systems.

\subsection{Real-world Performance}
\textbf{End-to-end performance.}
\begin{figure*}[t]
    \begin{subfigure}{0.24\textwidth}
        \centering
        \includegraphics[width=\textwidth]{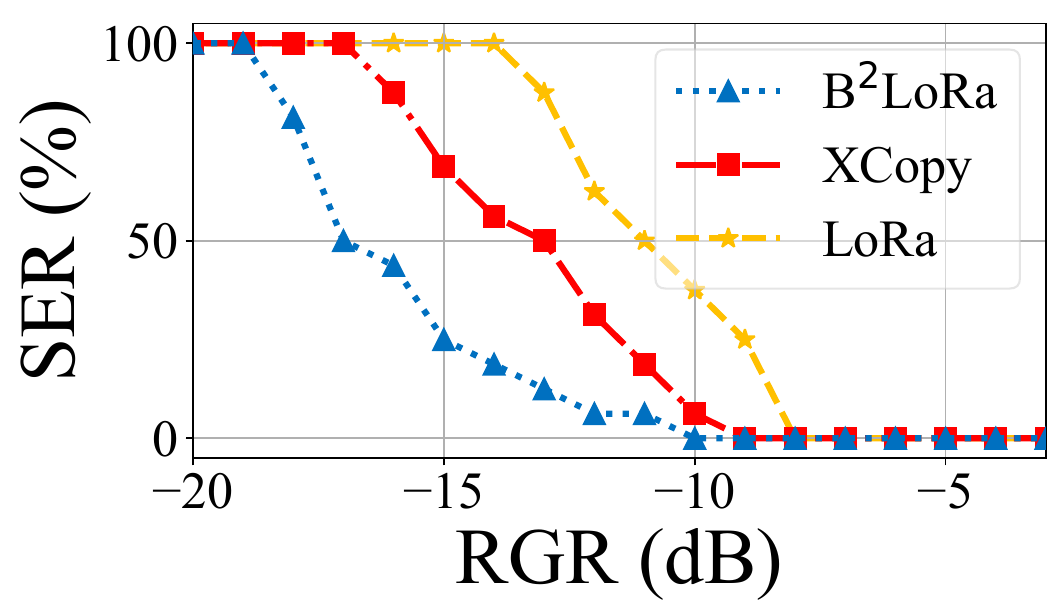}
        \vspace{-10mm}
        \captionsetup{justification=raggedright, singlelinecheck=false}
        \caption{}
        \label{fig: SER_SNR}
    \end{subfigure}
    \begin{subfigure}{0.24\textwidth}
        \centering
        \includegraphics[width=\textwidth]{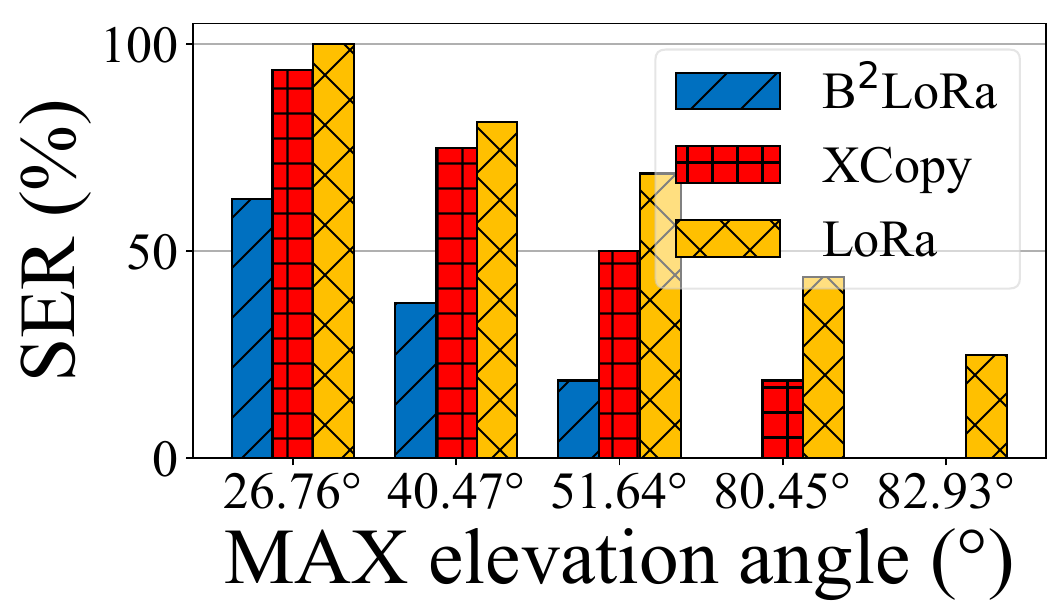}
        \vspace{-10mm}
        \captionsetup{justification=raggedright, singlelinecheck=false}
        \caption{}
        \label{fig: SER_EL}
    \end{subfigure}
    \begin{subfigure}{0.24\textwidth}
        \centering
        \includegraphics[width=\textwidth]{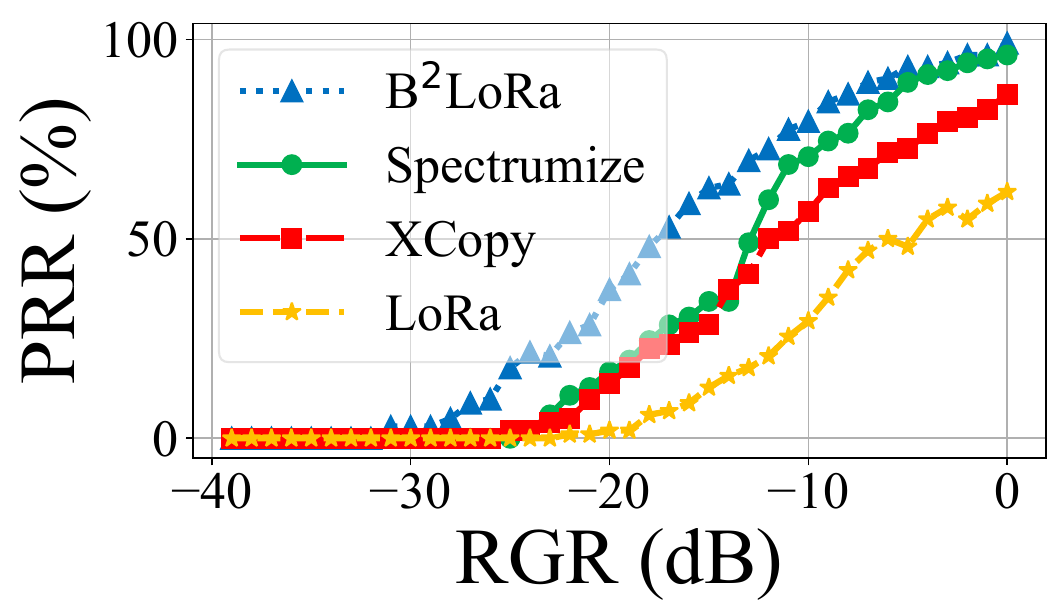}
        \vspace{-10mm}
        \captionsetup{justification=raggedright, singlelinecheck=false}
        \caption{}
        \label{fig: RR_SNR}
    \end{subfigure}
    \begin{subfigure}{0.24\textwidth}
        \centering
        \includegraphics[width=\textwidth]{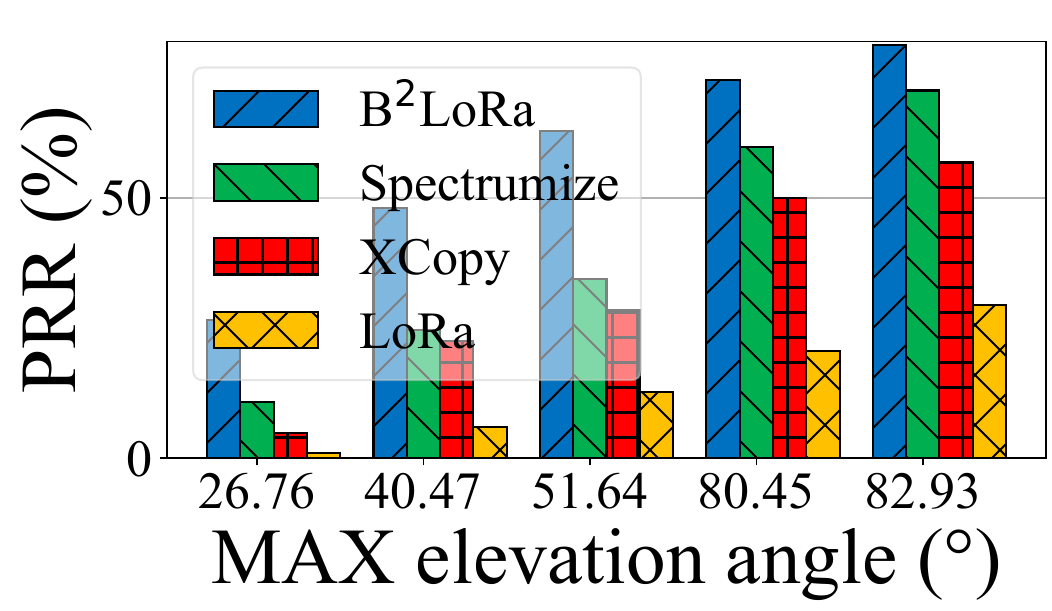}
        \vspace{-10mm}
        \captionsetup{justification=raggedright, singlelinecheck=false}
        \caption{}
    \label{fig: RR_EL_snr=-17}
    \end{subfigure}
    \caption{Real-world testbed: 
        (a) AVG SER vs. RGRs (during passes with MAX elevation angles greater than 80$\degree$); 
        (b) AVG SER during passes of different MAX elevation angles; 
        (c) AVG PRR vs. RGRs (during passes with MAX elevation angles greater than 80$\degree$);
        (d) AVG PRR during passes of different MAX elevation angles.
    }
\end{figure*}
Based on the real-world testbed, we evaluate the improvement of \BLoRa on packet decoding performance. In this experiment, both \BLoRa and XCopy combine multiple packets from re-transmissions into one, making the combined packet's SER reflective of their performance. For LoRa, we measure an SER for each re-transmission and select the lowest one to represent its performance.

Figure~\ref{fig: SER_SNR} presents the SERs of \BLoRa, LoRa, and XCopy at various RGRs when the satellite passes with a maximum elevation angle greater than $80\degree$. All schemes experience decoding errors below the RGR of \SI{-8}{\si{dB}}, but \BLoRa outperforms the others. For instance, at the RGR of \SI{-12}{\si{dB}}, \BLoRa's SER is about 5\%, which is 25\% lower than XCopy and 56\% lower than LoRa. Notably, at the RGR of \SI{-17}{\si{dB}}, while XCopy and LoRa fail to decode, \BLoRa maintains an SER of around 50\%, making it ideal for satellite-IoT systems.

Figure~\ref{fig: SER_EL} evaluates the SER performance during satellite passes of different maximum elevation angles when RGR is \SI{-9}{\si{dB}}. We observe that \BLoRa consistently performs the best. Notably, as the maximum elevation angle increases, the SER decreases. This is because when a satellite passes at a higher elevation angle relative to the receiver, the distance decreases, leading to less signal loss and increased SNRs.

\textbf{Packet detection.}
\begin{table}[t]
    \centering
    \caption{Detection capability gain relative to LoRa.}
    \vspace{-2mm}
    \resizebox{\linewidth}{!}{
        \begin{tabular}{c|ccccc}
            \toprule
            \textbf{Elevation angle} & 26.76° & 40.47° & 51.64° & 80.45° & 82.93° \\ 
            \hline
            \rowcolor{gray!10}\textbf{\BLoRa} & +9 dB & +9 dB & +9 dB & +9 dB & +9 dB \\
            \textbf{XCopy} & +3 dB & +3 dB & +3 dB & +4 dB & +4 dB \\
            \textbf{Spectrumize} & +4 dB & +4 dB & +5 dB & +5 dB & +6 dB \\
            \bottomrule
        \end{tabular}
    }
    \label{tab: detection_snr_boost}
    \vspace{-4mm}
\end{table}
In this experiment, we explore the packet detection performance of \BLoRa and compare it with Spectrumize, XCopy, and LoRa. As Spectrumize’s detection performance depends on the length of the virtual packet train, we iterate all lengths to yield its best performance.

Figure~\ref{fig: RR_SNR} shows the PRRs of the four schemes under various RGRs when the maximum elevation angle is greater than $80\degree$. Results indicate that when the RGR drops below \SI{-7}{\si{dB}}, \BLoRa's packet detection performance begins to surpass that of Spectrumize significantly and remains superior to both LoRa and XCopy. For instance, when RGR = \SI{-11}{\si{dB}}, \BLoRa's PRR is approximately 77\%, which is 9\% higher than Spectrumize, 26\% higher than XCopy, and 52\% higher than LoRa. It can be observed that \BLoRa's detection performance is less sensitive to RGR changes than Spectrumize. This is because Spectrumize relies on the ideal assumption of a constant interval between re-transmission arrivals, whereas \BLoRa can adapt to arrival time variations.

Table~\ref{tab: detection_snr_boost} reports the detection capability gain of \BLoRa, XCopy, and Spectrumize relative to standard LoRa at various maximum elevation angles. The gain is calculated as the reduction in the RGR threshold of detection. It is observed that \BLoRa consistently provides the most stable and significant gain of \SI{9}{\si{dB}} across various satellite passes.

Figure~\ref{fig: RR_EL_snr=-17} further evaluates the PRR performance of \BLoRa, Spectrumize, XCopy, and LoRa with RGR = \SI{-9}{\si{dB}} at various maximum elevation angles. \BLoRa consistently outperforms the other schemes, achieving an average PRR that is 17\% higher than Spectrumize, 25\% higher than XCopy, and 40\% higher than LoRa. PRR increases with rising maximum elevation angles, as expected, due to shorter link distances.

\textbf{Packet combination. }
\begin{figure}[t]
    \begin{subfigure}{0.23\textwidth}
        \centering
        \includegraphics[width=\textwidth]{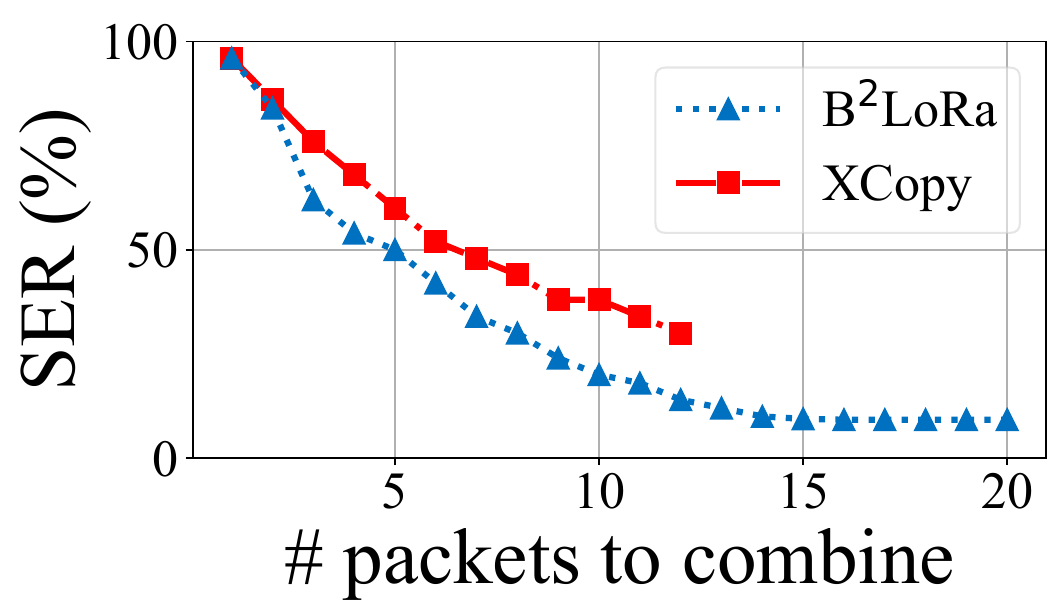}
        \captionsetup{justification=raggedright, singlelinecheck=false}
        \vspace{-9mm}
        \caption{}
        \label{fig: SER_retran}
    \end{subfigure}
        \begin{subfigure}{0.23\textwidth}
        \centering
        \includegraphics[width=\textwidth]{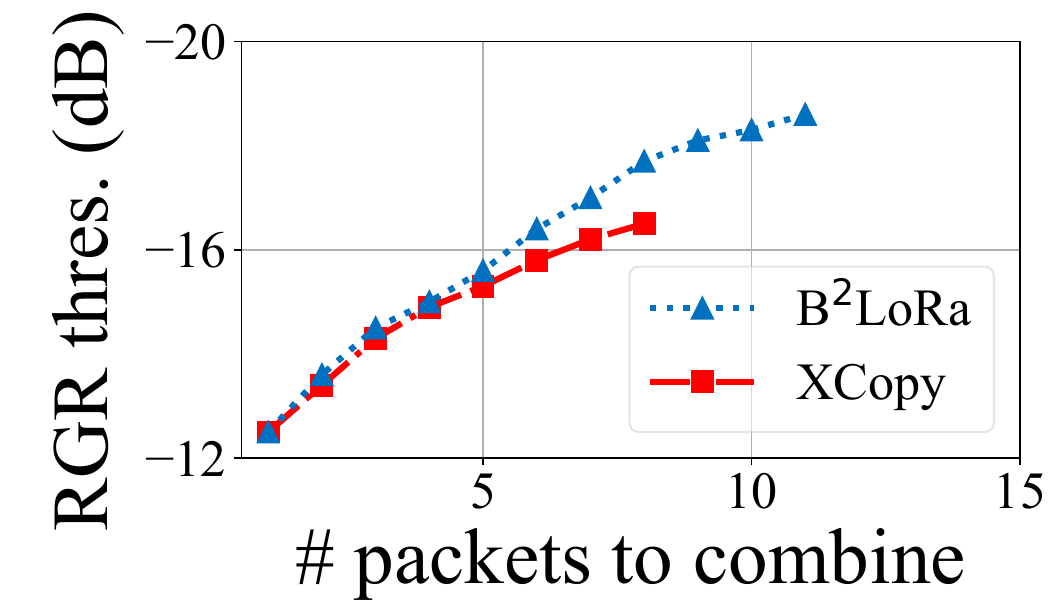}
        \captionsetup{justification=raggedright, singlelinecheck=false}
        \vspace{-9mm}
        \caption{}
        \label{fig: RGR_threshold}
    \end{subfigure}
    \caption{
        Real-world testbed:
        (a) SER vs. re-transmission times (measured at RGR = -12 dB during an 82.93$\degree$ pass);
        (b) RGR threshold of decoding vs. re-transmission times.
    }
    \vspace{-5mm}
\end{figure}
In this experiment, we examine the packet combination performance of \BLoRa and compare it with XCopy, which also performs packet combination. We set up the link during the satellite pass at a maximum elevation angle of $82.93\degree$. For both \BLoRa and XCopy, we configure the number of packets to combine from 1 to the maximum number of packets detected by each method.

Figure~\ref{fig: SER_retran} shows the SER performance of \BLoRa and XCopy as the number of packets to combine varies. The results demonstrate that with the same number of packets to combine, \BLoRa consistently achieves lower SER than XCopy. Notably, as the number of packets to combine increases, \BLoRa decreases more rapidly than XCopy in SER. This is attributed to \BLoRa's ability to compensate for the Doppler shift, which makes the benefits of a single combined packet more significant than those in XCopy.
Additionally, due to XCopy's weaker packet detection capabilities, it detects fewer packets during a satellite pass. Consequently, the upper limit on the number of packets to combine is lower for XCopy, as depicted in Figure~\ref{fig: SER_retran}, where the data points for XCopy are fewer than those for \BLoRa.

Figure~\ref{fig: RGR_threshold} shows the RGR threshold of decoding based on \BLoRa and XCopy with the different numbers of packets to combine. Notably, \BLoRa can decode at lower SNRs than XCopy when combining the same number of packets. Due to the weaker detection capability of XCopy, it detects only 7 packets for combining in this experiment. At this point, the decoding RGR threshold of \BLoRa is approximately \SI{2}{\si{dB}} lower than that of XCopy, indicating better noise immunity.

\textbf{Service range.}
\begin{figure}[t]
    \begin{subfigure}{0.21\textwidth}
        \centering
        \includegraphics[width=\textwidth]{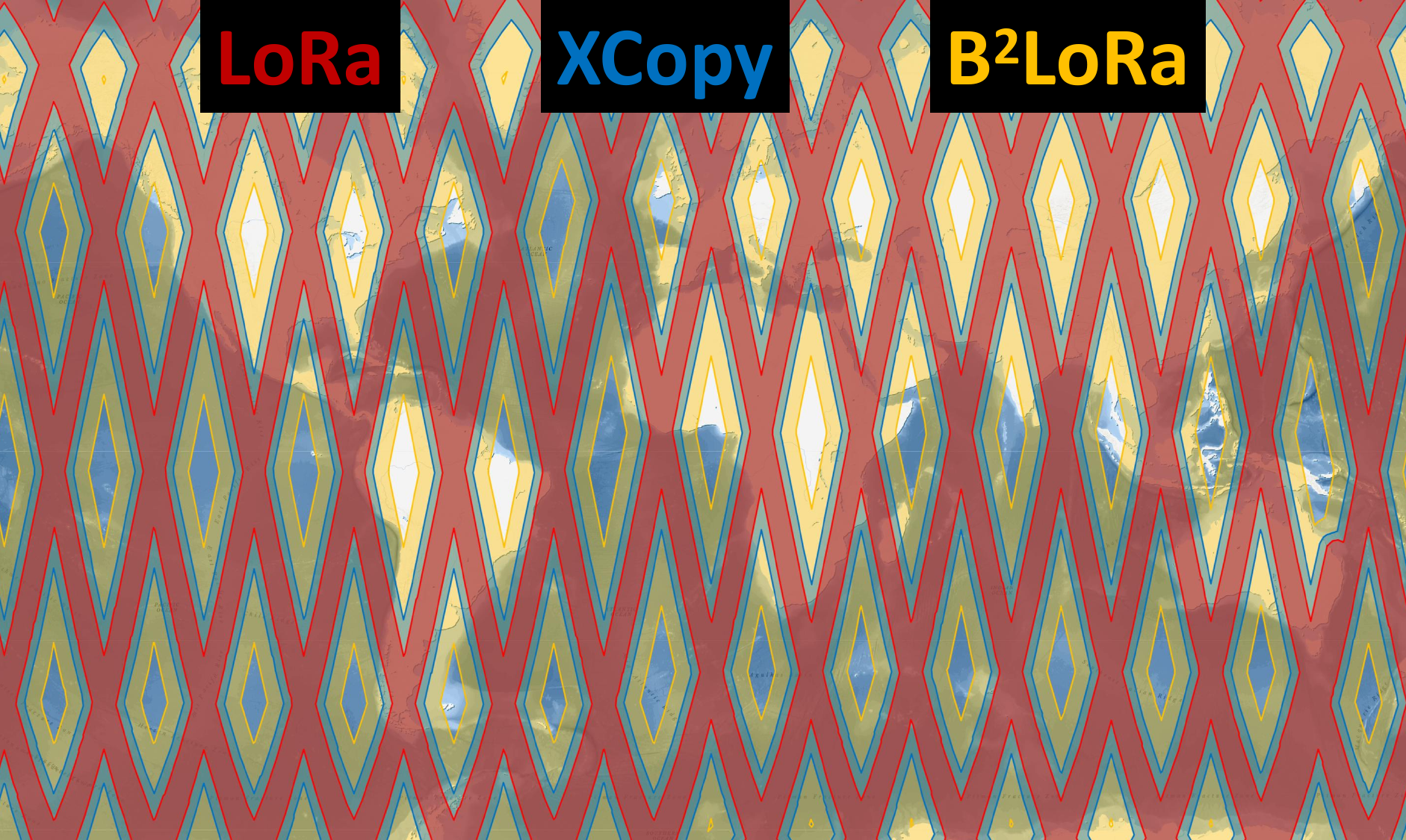}
        \vspace{-5.5mm}
        \caption{}
        \label{fig: service_coverage}
    \end{subfigure}
    \begin{subfigure}{0.23\textwidth}
        \centering
        \includegraphics[width=\textwidth]{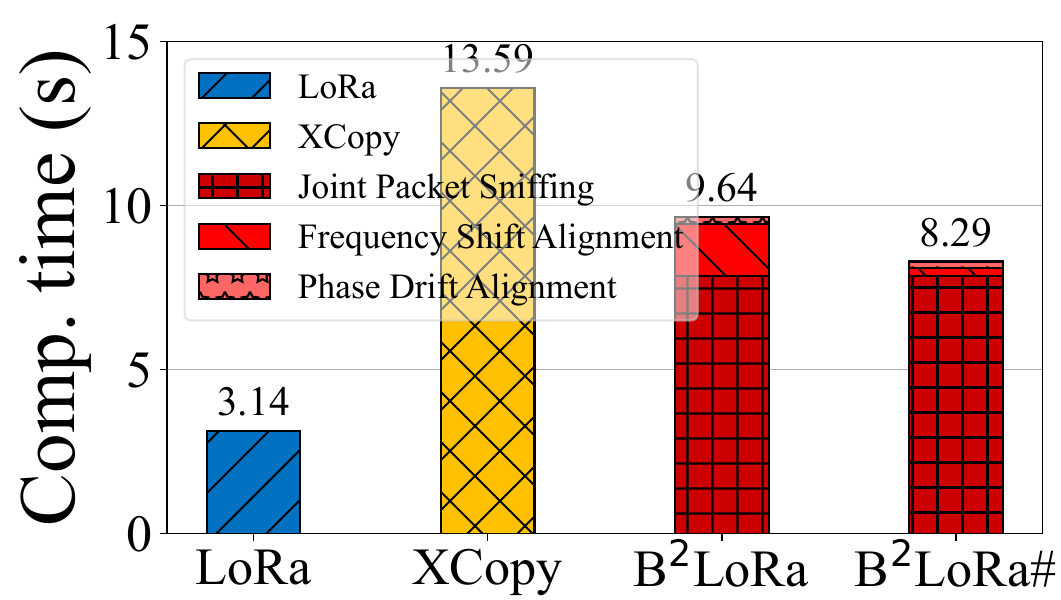}
        \vspace{-5.5mm}
        \caption{}
        \label{fig: overhead}
    \end{subfigure}
    \caption{
        (a) Our satellite's service range in one day adopting different methods;
        (b) AVG computational time for one pass ($\text{B}^2$LoRa\# replaces conjugate multiplication with orbital calculation for DFS alignment).
    }
\end{figure}
Figure~\ref{fig: service_coverage} shows the service range of \BLoRa, XCopy, and LoRa. \BLoRa offers significantly broader coverage, with a radius of $\approx$\SI{900}{\si{km}}, compared to XCopy ($\approx$\SI{500}{\si{km}}) and LoRa ($\approx$\SI{300}{\si{km}}). This indicates that \BLoRa can better make use of low-elevation-angle passes, thereby reducing the number of satellites required for global service.

\textbf{Overhead. }
Computational overhead is evaluated by processing data during a satellite pass ($\approx$12 min) on a laptop with different schemes. The results show time consumption ratios of 9.21:8.67:1 for \BLoRa, XCopy, and LoRa. Two insights are derived: 
(1) The main overhead for \BLoRa comes from FFTs, with the workload for joint packet sniffing increasing tenfold due to a $10.25\times$ window length compared to dechirping. 
(2) Frequency/phase alignment incurs relatively small overhead since their FFT computations require only a few iterations compared to the sliding FFT during detection. 
Notably, the empirical time consumption ratio between \BLoRa and LoRa is actually 2.64:1 (Figure~\ref{fig: overhead}) for the following reasons. 

(1) \BLoRa does not need to process the entire 12 minutes of data comprehensively. Across various passes evaluated, \BLoRa only needs to process on average 222 seconds of I/Q data (i.e., combining 8 packets) to successfully decode at least one packet, without needing to process the remaining 9 minutes of data. In contrast, due to its weaker detection capability and smaller gain per packet combined, XCopy requires processing a longer portion of data. LoRa even needs to process the complete 12-minute data. 

(2) $\text{B}^2$LoRa’s rotating conjugate multiplication can be employed only during the cold-start phase of IoT devices. Upon successfully receiving the first TLE beacon, the satellite orbit is known, \BLoRa uses orbital calculations instead for active Doppler shift compensation, considerably reducing computational overhead. This Doppler-compensation-optimized scheme is measured and denoted as $\text{B}^2$LoRa\# in Figure~\ref{fig: overhead}.

Regarding storage overhead, \BLoRa collects about 3 GB of samples for blind coherent combining, which is higher than conventional schemes. Using a COTS 16 GB SD card results in an approximate cost increase of 7.5 USD.

\textbf{Ablation study.}
\begin{figure}[t]
    \begin{subfigure}{0.23\textwidth}
        \centering
        \includegraphics[width=\textwidth]{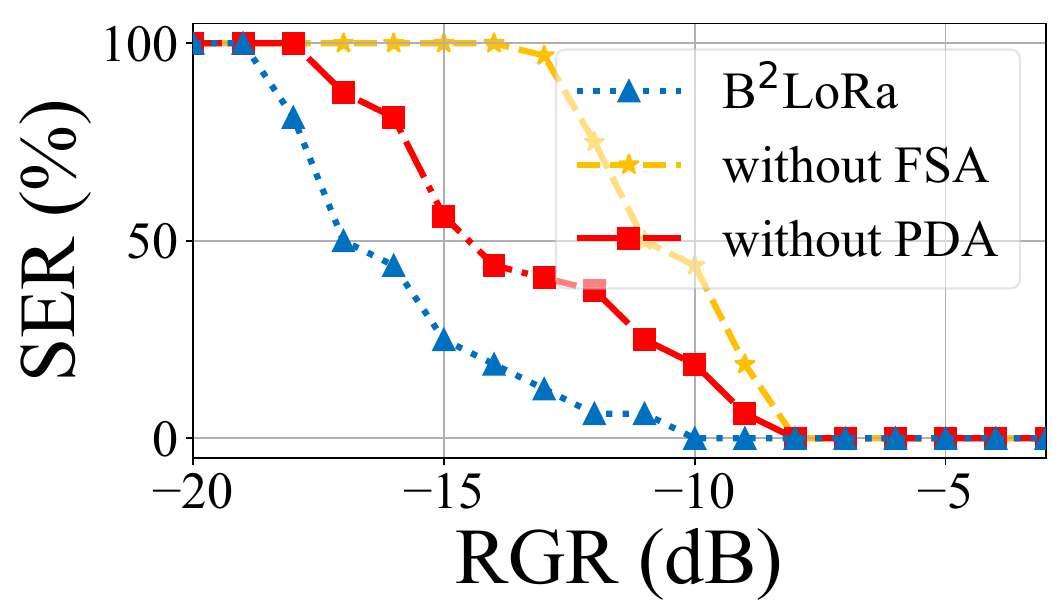}
        \vspace{-10mm}
        \captionsetup{justification=raggedright, singlelinecheck=false}
        \caption{}
        \label{fig: SER_SNR_ablation}
    \end{subfigure}
    \begin{subfigure}{0.23\textwidth}
        \centering
        \includegraphics[width=\textwidth]{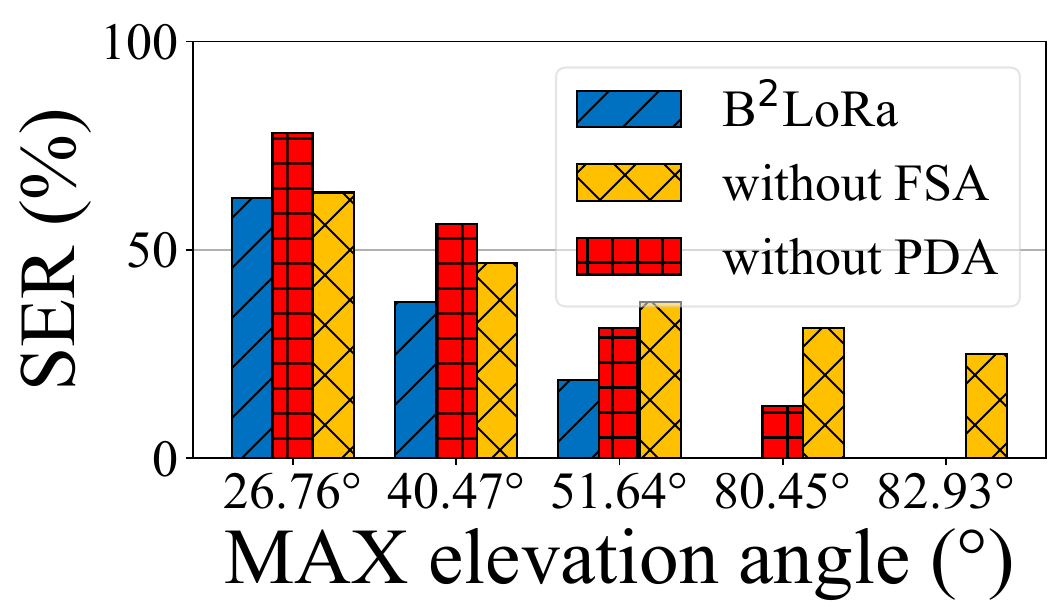}
        \vspace{-10mm}
        \captionsetup{justification=raggedright, singlelinecheck=false}
        \caption{}
        \label{fig: SER_EL_snr=-9_ablation}
    \end{subfigure}
    \caption{Ablation study:
        (a) AVG SER vs. RGRs (during passes with MAX elevation angles greater than 80$\degree$);
        (b) AVG SER during various satellite passes.
    }
    \vspace{-5mm}
\end{figure}
To assess how each component influences the end-to-end performance, we perform an ablation study with three scenarios: \BLoRa, \BLoRa without intra-packet Doppler Frequency Shift Alignment (FSA), and \BLoRa without inter-packet Phase Drift Alignment (PDA). Performance is evaluated by the SER of the combined packets.

Figure~\ref{fig: SER_SNR_ablation} presents the SERs across the three scenarios at various RGRs measured when the satellite's maximum elevation angle exceeds 80°. Two insights can be observed: (1) The gain provided by PDA remains relatively stable across different noise intensities; (2) When the noise is sufficiently strong (e.g., RGR < -12 dB), FSA becomes crucial, contributing nearly 5 dB of gain to the decoding performance of \BLoRa. Figure~\ref{fig: SER_EL_snr=-9_ablation} evaluates the SER performance at an RGR of -9 dB during various passes at different elevation angles. It can be noticed that (1) the gain provided by FSA is significant but gradually decreases as the elevation angle lowers, and (2) the gain provided by PDA remains relatively consistent.

To summarize, the intra-packet Doppler frequency shift, which terrestrial systems often overlook, is vital to handle in satellite-IoT systems. Additionally, the initial phase search method proves to be cost-effective, requiring minimal computation while providing stable and considerable gains.

\subsection{SDR-based Benchmark}
\begin{figure*}[t]
    \centering
    \begin{subfigure}{0.24\textwidth}
        \centering
        \includegraphics[width=\textwidth]{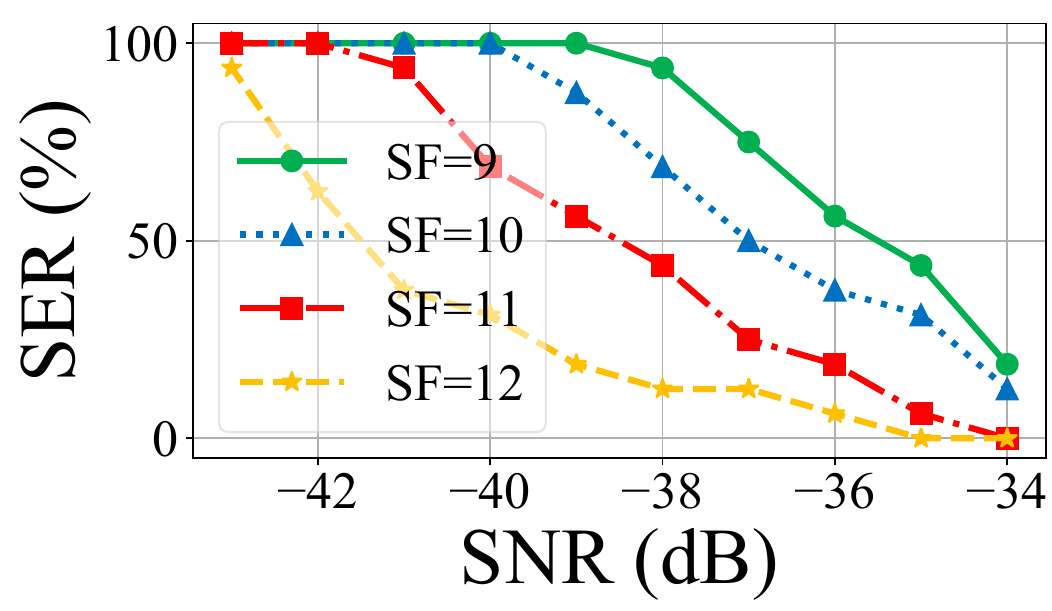}
        \vspace{-10mm}
        \captionsetup{justification=raggedright, singlelinecheck=false}
        \caption{}
        \label{fig: SER_SNR_Var_SF}
    \end{subfigure}
    \begin{subfigure}{0.24\textwidth}
        \centering
        \includegraphics[width=\textwidth]{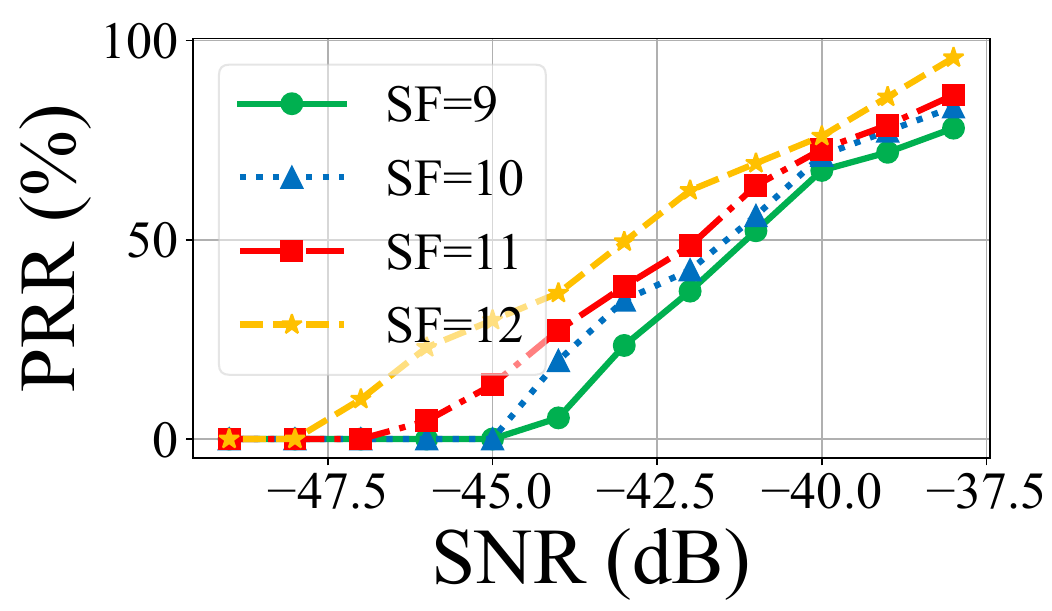}
        \vspace{-10mm}
        \captionsetup{justification=raggedright, singlelinecheck=false}
        \caption{}
        \label{fig: RR_SNR_Var_SF}
    \end{subfigure}
    \begin{subfigure}{0.24\textwidth}
        \centering
        \includegraphics[width=\textwidth]{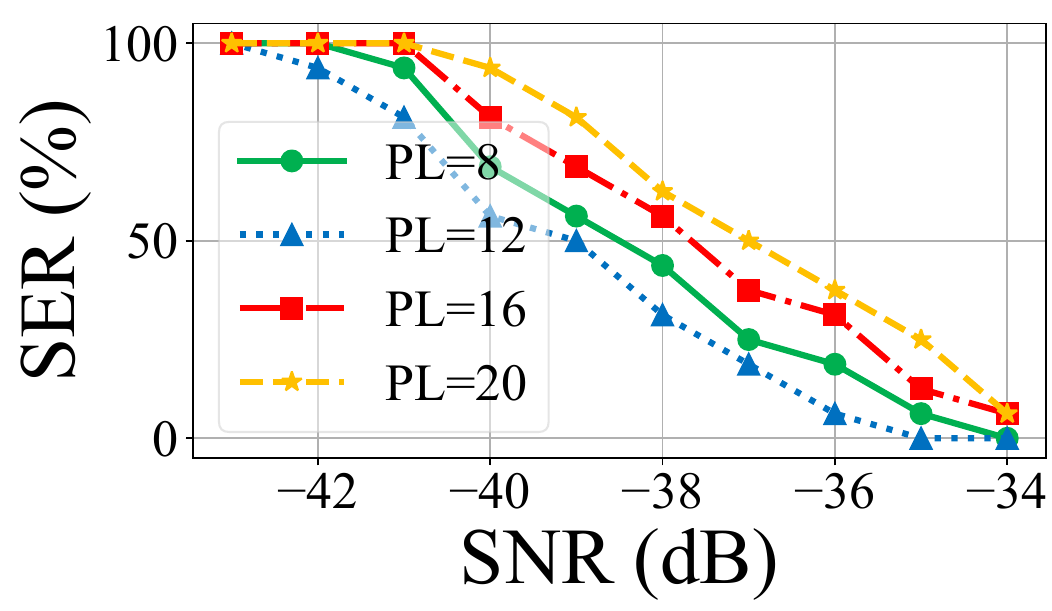}
        \vspace{-10mm}
        \captionsetup{justification=raggedright, singlelinecheck=false}
        \caption{}
        \label{fig: SER_SNR_Var_PL}
    \end{subfigure}
    \begin{subfigure}{0.24\textwidth}
        \centering
        \includegraphics[width=\textwidth]{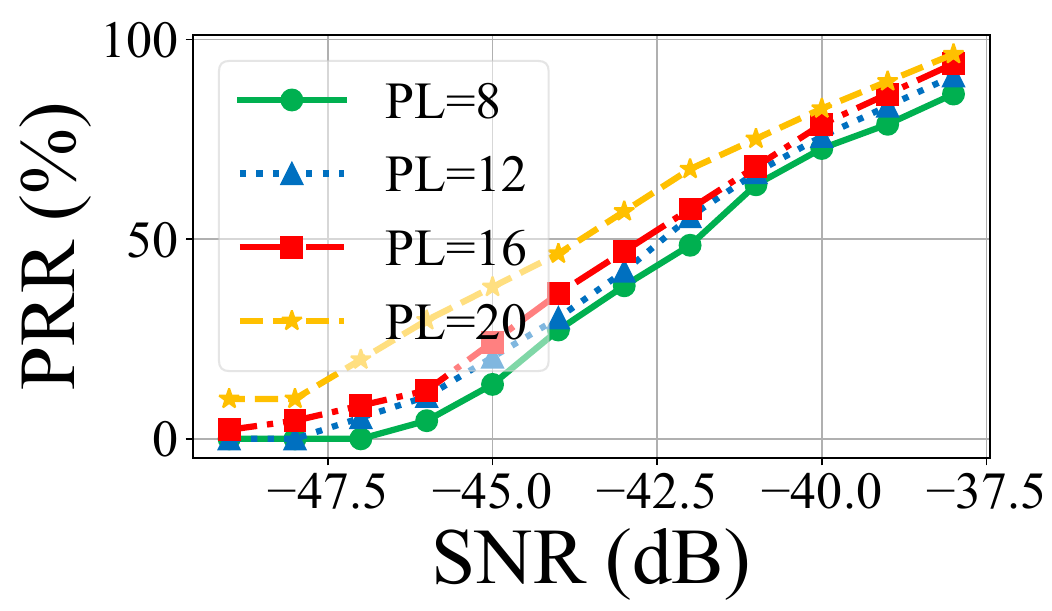}
        \vspace{-10mm}
        \captionsetup{justification=raggedright, singlelinecheck=false}
        \caption{}
        \label{fig: RR_SNR_Var_PL}
    \end{subfigure}
    \begin{subfigure}{0.24\textwidth}
        \centering
        \includegraphics[width=\textwidth]{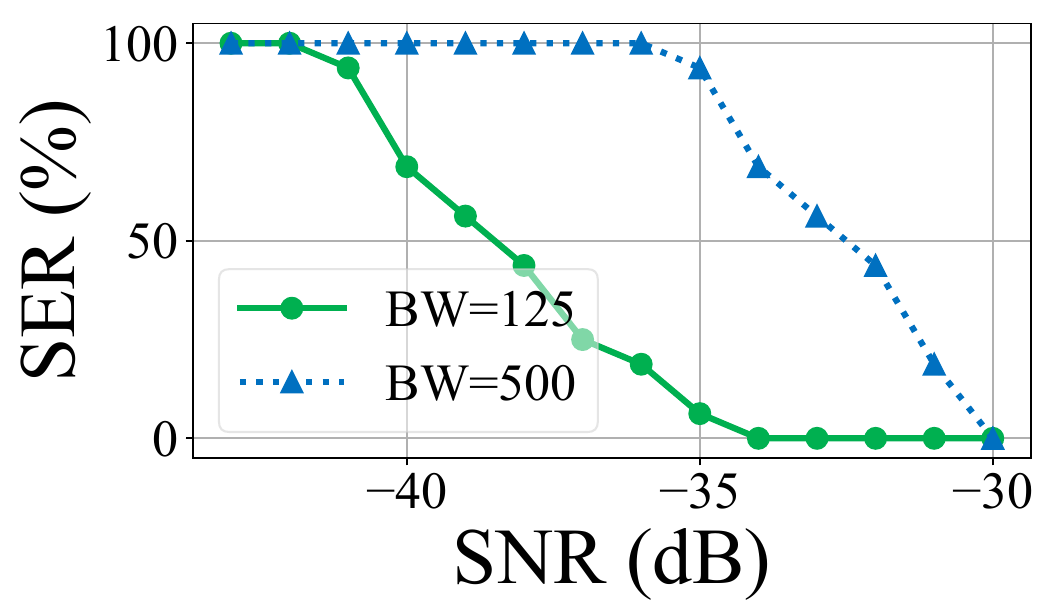}
        \vspace{-10mm}
        \captionsetup{justification=raggedright, singlelinecheck=false}
        \caption{}
        \label{fig: SER_SNR_Var_BW}
    \end{subfigure}
    \begin{subfigure}{0.24\textwidth}
        \centering
        \includegraphics[width=\textwidth]{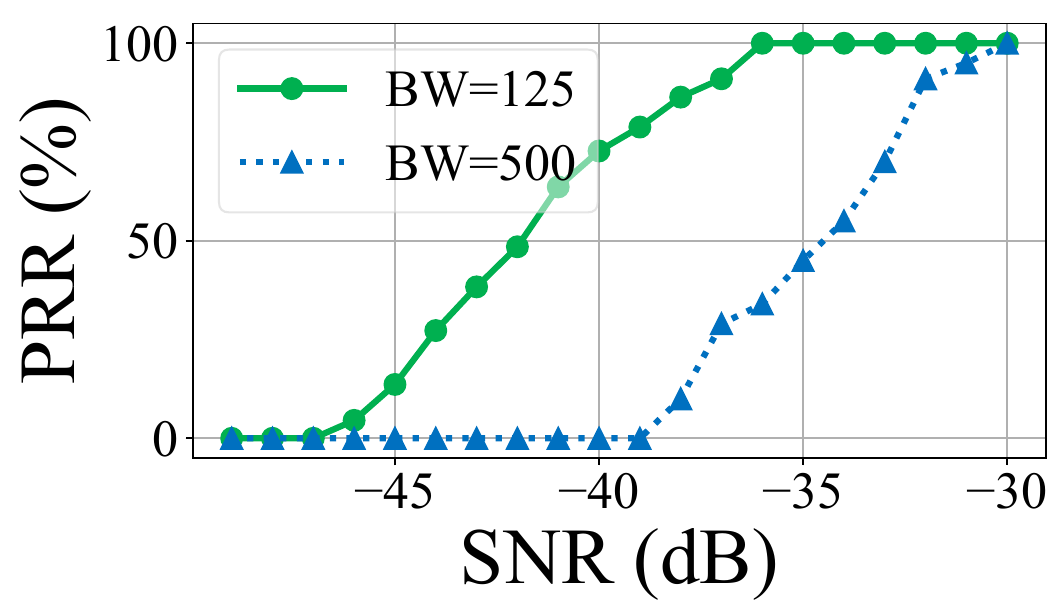}
        \vspace{-10mm}
        \captionsetup{justification=raggedright, singlelinecheck=false}
        \caption{}
        \label{fig: RR_SNR_Var_BW}
    \end{subfigure}
    \begin{subfigure}{0.24\textwidth}
        \centering
        \includegraphics[width=\textwidth]{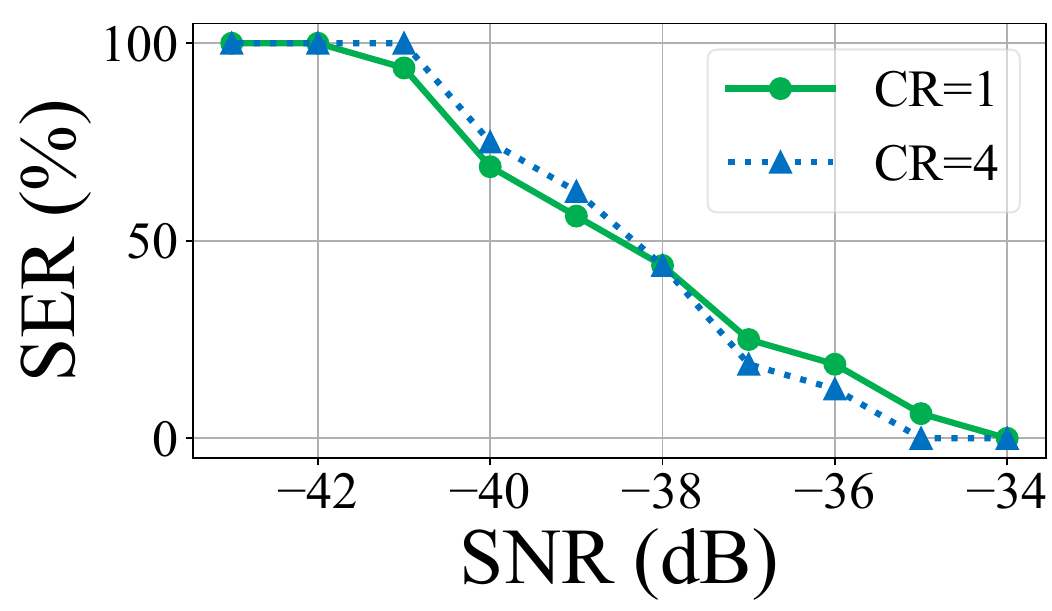}
        \vspace{-10mm}
        \captionsetup{justification=raggedright, singlelinecheck=false}
        \caption{}
        \label{fig: SER_SNR_Var_CR}
    \end{subfigure}
    \begin{subfigure}{0.24\textwidth}
        \centering
        \includegraphics[width=\textwidth]{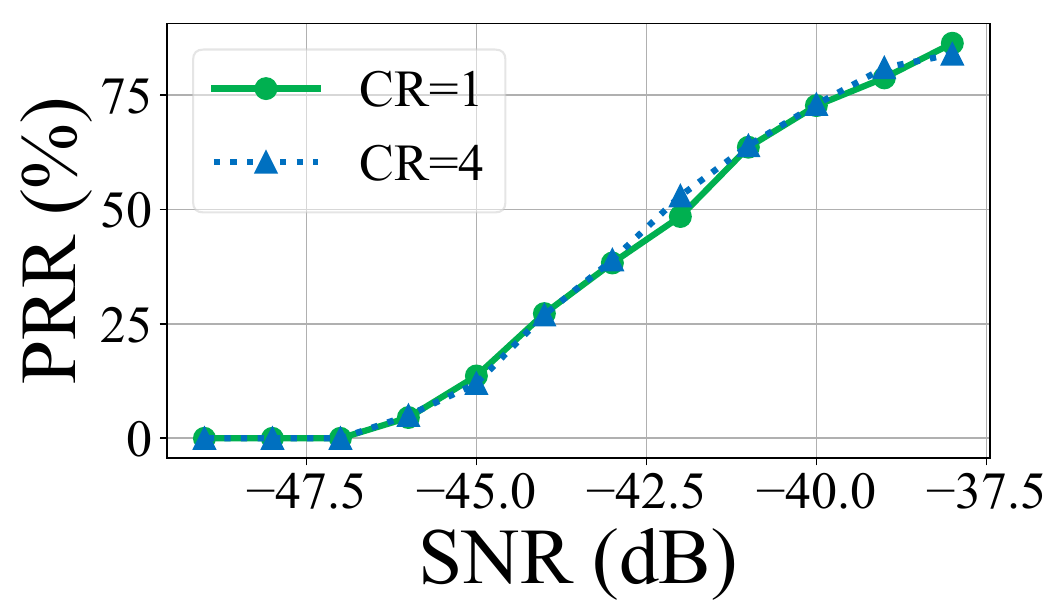}
        \vspace{-10mm}
        \captionsetup{justification=raggedright, singlelinecheck=false}
        \caption{}
        \label{fig: RR_SNR_Var_CR}
    \end{subfigure}
    \caption{SDR-based benchmark:
        (a,b) Impact of Spreading Factors (SFs);
        (c,d) Impact of Preamble Lengths (PLs);
        (e,f) Impact of Bandwidths (BWs); 
        (g,h) Impact of Coding Rates (CRs).
    }
\end{figure*}
The above experiments are conducted on a real-world testbed with fixed LoRa PHY settings due to legal constraints. We also examine the impact of the Spreading Factor, Preamble Length, Coding Rate, and Bandwidth on \BLoRa's decoding and detection performance on an SDR-based testbed, which allows precise signal strength settings. The highest SNR during the satellite pass is used as the signal strength metric.

\textbf{Spreading Factor (SF).} 
We set PL=8, BW=\SI{125}{\si{kHz}}, and CR=1, with the SF varying from 9 to 12. Figures~\ref{fig: SER_SNR_Var_SF} and~\ref{fig: RR_SNR_Var_SF} show the SER and PRR performance under different SFs. It can be observed that larger SFs lead to improved SER and PRR. Specifically, when SF=12, the SER is about 51\% lower, and the PRR is approximately 17\% higher compared to SF=9. This superior performance results from a larger SF's ability to accumulate more energy and produce higher FFT peaks, enhancing noise resistance during detection and decoding.

\textbf{Preamble Length (PL).} 
We set SF=11, BW=\SI{125}{\si{kHz}}, and CR=1, with the PL varying from 8 to 20. Figure~\ref{fig: SER_SNR_Var_PL} shows the SER performance under different PLs. Notably, PLs beyond 12 do not reduce SER, with PL=16 showing about 8\% higher SER than PL=8. In our view, this occurs because longer PLs excessively boost $\text{B}^2$LoRa's detection capability, resulting in the detection of some packets with insufficient power. \textit{In such cases, a packet deeply buried within noise brings a more negative effect than positive in the coherent combining process.} Figure~\ref{fig: RR_SNR_Var_PL} presents PRRs across PLs. Results show that extending PL can enhance PRR. For instance, PL=20 yields approximately 15\% higher PRR than PL=8. This can be attributed to greater energy accumulation in chaining-dechirp, thus improving the detection capability.

\textbf{Bandwidth (BW).}
We set PL=8, SF=11, and CR=1, with BWs of \SI{125}{\si{kHz}} and \SI{500}{\si{kHz}}. Figures~\ref{fig: SER_SNR_Var_BW} and~\ref{fig: RR_SNR_Var_BW} illustrate that lower BW improves \BLoRa's detection and decoding. Specifically, at \SI{500}{\si{kHz}}, the SER is about 41\% higher, and the PRR is roughly 35\% lower compared to 125 kHz. This is because reducing BW increases chirp length, benefiting chaining-dechirp performance with more energy.

\textbf{Coding Rate (CR).}
We set PL=8, SF=11, and BW=\SI{125}{\si{kHz}}, with CRs of 1 and 4. Figures~\ref{fig: SER_SNR_Var_CR} and~\ref{fig: RR_SNR_Var_CR} show the SER and PRR performance under different CRs. 
The results indicate that CR minimally impacts performance, as it only affects payload length without altering the chirp physical structure, keeping chaining-dechirp performance consistent. Thus, the number of packets for coherent combining and resulting SNRs remain unchanged, and SERs stay consistent.
\section{Related Works}
In this section, we discuss prior research on satellite-IoT communication and LoRa link performance.

\textbf{Satellite-IoT communication.}
Research on satellite-IoT communication primarily targets LoRa~\cite{colavolpe2019reception, wu2019research, s22051830, SALSA, fraire2022space} and NB-IoT~\cite{cluzel20183gpp, sciddurlo2021looking, kodheli2021nb, mannoni2021nb}. LoRa satellites enable applications in remote regions, like offshore wind farm monitoring~\cite{ullah2021enabling} and ocean surveillance~\cite{fernandez2020evaluation}. With long range, link performance improvements are researched. For instance, PMSat~\cite{PMSat} uses a passive metasurface for beamforming to enhance SNRs. However, it cannot be applied to \BLoRa due to the high hardware and computational costs. Meanwhile, the latest work, Spectrumize~\cite{Spectrumize}, relates closely to \BLoRa, both aiming to improve link performance in LoRa-based satellte-IoT systems. Spectrumize focuses on ground stations and requires prior knowledge of the satellites' ephemeris for compensating Doppler shifts to enhance packet detection capabilities. This requirement is infeasible for IoT devices since they lack information about satellites' real-time locations. Furthermore, Spectrumize does not address the challenge of demodulating and decoding weak packets after detecting them.

\textbf{LoRa link performance.}
Substantial works aim to improve the link performance of LoRa communication~\cite{Choir, mLoRa, FTrack, OCT, NScale, CIC, Pyramid, PCube, XGate, Charm, MALoRa, XCopy, Spectrumize, fu2021real, sandoval2019optimizing, dix2018lora, DyLoRa, Falcon}, which can be categorized into three groups. The first category includes studies that enhance LoRa links through parameter optimization~\cite{fu2021real, sandoval2019optimizing, dix2018lora} or dynamic selection~\cite{DyLoRa}. The second category addresses the collision problem in LoRa communications, with most studies focusing on terrestrial networks. These studies exploit unique time, frequency, and phase offsets of packets as fingerprints to cancel out collisions, thus improving link performance~\cite{Choir, mLoRa, FTrack, OCT, NScale, CIC, Pyramid, PCube, XGate}. Notably, Spectrumize~\cite{Spectrumize} targets satellite-IoT systems and addresses packet collisions via Doppler frequency shifts, where different shifts act as unique fingerprints corresponding to the orbital passes of various satellites. Finally, some research concentrates on overcoming low link budgets in terrestrial networks by coherently combining multiple packets from dual transmitters~\cite {Falcon}, various LoRa gateways~\cite{Charm, MALoRa}, or re-transmissions~\cite{XCopy}. However, in satellite-IoT systems characterized by ultra-low link budgets and varying Doppler frequency shifts, existing designs encounter two issues: (1) difficulty in detecting sufficient LoRa packets and (2) neglect of Doppler frequency shifts, which prevents coherent combining and results in significant loss of gain.
\section{Conclusion}
In this paper, we present a blind coherent combining design named \BLoRa to boost LoRa link performance in satellite-IoT systems. 
\BLoRa leverages the repeated broadcasting mechanism inherent in satellite-IoT systems for coherent combining, all while maintaining low power and cost without needing channel feedback. To achieve coherent combining at a fine granularity, \BLoRa incorporates three designs: joint packet sniffing, frequency shift alignment, and phase drift mitigation to address the challenges of frequent packet loss and Doppler frequency shifts.
Our extensive real-world experiments demonstrate that \BLoRa significantly boosts SNRs for IoT devices and effectively enhances link reliability.

\section{Acknowledgement}
We thank our shepherd, Renjie Zhao, and the anonymous reviewers for their insightful comments. This work is supported in part by the National Science and Technology Innovation 2030 Major Project of China (No. 2022ZD0208603), the Shanghai Central Guidance Local Science and Technology Development Fund Project (No. YDZX20253100004000), the NSFC (No. W2412089, 62172277), and the Microsate Project (No. MICR-KY-2024010028). Xiong Wang and Linghe Kong are the corresponding authors.

\bibliographystyle{unsrt}
\bibliography{sigconf}

\end{document}